\documentclass[fleqn,usenatbib]{mnras}

\usepackage{newtxtext,newtxmath}
\usepackage[T1]{fontenc}

\DeclareRobustCommand{\VAN}[3]{#2}
\let\VANthebibliography\thebibliography
\def\thebibliography{\DeclareRobustCommand{\VAN}[3]{##3}\VANthebibliography}

\usepackage{graphicx}	% Including figure files
\usepackage{amsmath}	% Advanced maths commands

\usepackage{amssymb}	% Extra maths symbols
\usepackage{bm}

\title[Dynamical cluster masses from photometric surveys]{Dynamical cluster masses from photometric surveys}

\author[O. Contigiani et al.]{Omar Contigiani,$^{1}$\thanks{E-mail: contigiani@cita.utoronto.ca}, Henk Hoekstra$^{2}$, Margot M. Brouwer,$^{4,5}$
Andrej Dvornik,$^{6}$ \newauthor Maria Cristina Fortuna,$^{1}$ Cristóbal Sifón,$^{7}$ Ziang Yan,$^{6}$ Mohammadjavad Vakili$^{1}$
\\
$^{1}$Canadian Institute for Theoretical Astrophysics, University of Toronto \\
$^{2}$Leiden Observatory, Leiden University \\
$^{4}$Kapteyn Astronomical Institute, University of Groningen \\
$^{5}$Institute for Theoretical Physics, University of Amsterdam \\
$^{6}$Ruhr-University Bochum, Astronomical Institute, German Centre for Cosmological Lensing, Universitätsstr. 150, 44801 Bochum, Germany \\
$^{7}$Instituto de F\'isica, Pontificia Universidad Cat\'olica de Valpara\'iso, Casilla 4059, Valpara\'iso, Chile \\
}

\date{Accepted XXX. Received YYY; in original form ZZZ}

\pubyear{2021}

\begin{document}
\label{firstpage}
\pagerange{\pageref{firstpage}--\pageref{lastpage}}
\maketitle

\begin{abstract}
The masses of galaxy clusters can be measured using  data obtained exclusively from wide photometric surveys in one of two ways: directly from the amplitude of the weak lensing signal or, indirectly, through the use of scaling relations calibrated using binned lensing measurements. In this paper, we build on a recently proposed idea and implement an alternative method based on the radial profile of the satellite distribution. This technique relies on splashback, a feature associated with the apocenter of recently accreted galaxies that offers a clear window into the phase-space structure of clusters without the use of velocity information.  We carry out this dynamical measurement using the stacked satellite distribution around a sample of luminous red galaxies in the fourth data release of the Kilo-Degree Survey and validate our results using abundance-matching and lensing masses. To illustrate the power of this measurement, we combine dynamical and lensing mass estimates to robustly constrain scalar-tensor theories of gravity at cluster scales. Our results exclude departures from General Relativity of order unity. We conclude the paper by discussing the implications for future data sets. Because splashback mass measurements scale only with the survey volume, stage-IV photometric surveys are well-positioned to use splashback to provide high-redshift cluster masses.
\end{abstract}


%%%%%%%%%%%%%%%%%%%%%%%%%%%%%%%%%%%%%%%%%%%%%%%%%%

%%%%%%%%%%%%%%%%% BODY OF PAPER %%%%%%%%%%%%%%%%%%

\section{Introduction}
The majority of ordinary matter, a.k.a. baryonic matter, is trapped inside the potential wells of the large-scale structure of the Universe. The main constituent of this invisible scaffolding is dark matter, and its fully collapsed overdensities, known as haloes, contain most of the mass in the Universe. These structures are not isolated, and the process of structure formation is known to be hierarchical \citep{1974ApJ...187..425P}. In simple terms, this means that smaller haloes become subhaloes after they are accreted onto larger structures. Unsurprisingly, baryonic matter also follows this process, resulting in today's clusters of galaxies. Due to their joint evolution, a tight relationship exists between the luminosity of a galaxy and the mass of the dark matter halo it inhabits.
These galaxy clusters are associated with the largest haloes in the Universe and they are still accreting matter from the surrounding environment, i.e. they are not fully virialized yet. 

Galaxies can be divided into two populations: red and blue \citep{2001AJ....122.1861S}. Whereas red galaxies derive their color from their aging stellar population, blue galaxies display active star formation, and young stars dominate their light. The exact mechanism behind quenching, i.e., the transition from star-forming to ``red and dead'', is still not fully understood \citep[see, e.g.,][]{2010MNRAS.402.1536S, 2015MNRAS.452.2879T}, but it is known to be connected to both baryonic feedback \citep[see, e.g.,][]{2008MNRAS.391..481S, 2010MNRAS.402.1536S} and interactions inside the dense cluster environment \citep[see, e.g.,][]{1980ApJ...237..692L, 1996Natur.379..613M, 2008MNRAS.387...79V}. An important consequence of this environmental dependence is the formation of a red sequence, i.e., a close relationship between the color and magnitude of red galaxies in clusters. By calibrating this red sequence as a function of redshift, it is possible to identify clusters in photometric surveys, even in the absence of precise spectroscopic redshifts \citep{2000AJ....120.2148G}. 

In recent years, splashback has been recognized as a feature located at the edge of galaxy clusters. The radius of this boundary, $r_\text{sp}$, is close to the apocenter of recently accreted material \citep[see, e.g.,][]{Adhikari_2014, Diemer_2017, Diemer_2017b} and it is associated with a sudden drop in matter density. This is because it naturally separates the single and multi-stream regions of galaxy clusters: orbiting material piles up inside this radius, while collapsing material located outside it is about to enter the cluster for the first time.

In simulations and observations, the distribution of red satellite galaxies and dark matter seem to trace this feature in the same fashion \citep{2021MNRAS.tmp.1404C, 2021MNRAS.504.4649O}, but a possible dependence on satellite properties is currently being explored \citep{2021arXiv210505914S, 2022arXiv220205277O}. In fact, in the context of galaxy evolution models, the mechanism behind this feature has been known under the name backsplash for almost two decades and has been previously explored both in observations and simulations \citep{2005MNRAS.356.1327G, 2011MNRAS.416.2882M}. Compared to these efforts, however, the recent interest in this feature is guided by theoretical and observational implications for the study of the large-scale structure of the Universe. 

Since haloes are perturbations on top of a background of constant density, their size can be quantified in terms of overdensity masses. For example, $M_\text{200m}$ is defined as the mass contained within a sphere of radius $r_\text{200m}$ such that the average density within it is $200$ times the average matter density of the Universe $\rho_\text{m}(z)$,
\begin{equation}
\label{eq:200m}
    M_\text{200m} = 200 \times \frac{4\pi}{3} \rho_\text{m}(z) r_\text{200m}^3.
\end{equation}
From a theoretical perspective, the splashback radius defines a more accurate cluster mass and sidesteps the issue of pseudo evolution due to an evolving $\rho_\text{m}(z)$ as a function of redshift $z$ \citep{2013ApJ...766...25D, More_2015}. Thanks to this property, this definition implies a universal mass function that is valid for a variety of cosmologies \citep{2020ApJ...903...87D}. Moreover, the shape of the matter profile around this feature can also be used to learn about structure formation, the nature of dark matter \citep{2020JCAP...02..024B} and dark energy \citep{Contigiani_2019}.

Observationally, one of the most noteworthy applications of the splashback feature is the study of quenching through the measurement of the spatial distribution of galaxy populations with different colors \citep{2020arXiv200811663A}. While notable, this was not the earliest result from the literature, and many other measurements preceded it. Published works can be divided into three groups: those based on targeted weak lensing observations of X-ray selected clusters \citep{2017ApJ...836..231U, 2019MNRAS.485..408C}, those based on the lensing signal and satellite distributions around SZ-selected clusters \citep[see, e.g., ][]{Shin_2019}, and those based on samples constructed with the help of cluster-finding algorithms applied to photometric surveys \citep[see, e.g.,][]{2016ApJ...825...39M, 2018ApJ...864...83C}. However, we note that in the case of the last group, the results are difficult to interpret because the splashback signal correlates with the parameters of the cluster detection method \citep{Busch_2017}. 

In this work, we implement an application of this feature based on \cite{2021MNRAS.tmp.1404C}. The location of the splashback radius is connected to halo mass, and its measurement from the distribution of cluster members can therefore lead to a mass estimate. Because this distribution can be measured without spectroscopy, this means that we can extract a dynamical mass purely from photometric data. To avoid the issues related to cluster-finding algorithms explained above, we studied the average distribution of faint galaxies around luminous red galaxies (LRGs) instead of the targets identified through overdensities of red galaxies. If we consider only passive evolution, the observed magnitude of the LRGs can be corrected to construct a sample with constant comoving density \citep{2016MNRAS.461.1431R,2019MNRAS.487.3715V}, and, by selecting the brightest among them, we expect to identify the central galaxies of groups and clusters. 

We present our analysis in Section~\ref{sec:profiles} and produce two estimates of the masses of the haloes hosting the LRGs in Section~\ref{sec:fit}. The first is based on the splashback feature measured in the distribution of faint galaxies, while the second is based on the amplitude of weak lensing measurements. After comparing these results with an alternative method in Section~\ref{sec:discussion}, we discuss our measurements in the context of modified models of gravity. We conclude by pointing out that, while we limit ourselves to redshifts $z<0.55$ here, the sample constructed in this manner has implications for the higher redshift range probed by future stage-IV photometric surveys \citep{2006astro.ph..9591A} such as
\emph{Euclid} \citep{laureijs2011euclid} and the Legacy Survey of Space and Time \citep[LSST,][]{2009arXiv0912.0201L}. 
Section~\ref{sec:future} discusses these complications in more detail and explores how this method can be used to complement the use of lensing to extract the masses of X-ray \citep{2019MNRAS.485..408C} or SZ selected clusters \citep{Shin_2019}.

Unless stated otherwise, we assume a cosmology based on the 2015 Planck data release \citep{Planck2015}. For cosmological calculations, we use the Python packages \textsc{astropy} \citep{Price-Whelan:2018hus} and \textsc{colossus} \citep{Diemer:2017bwl}. The symbols $R$ and $r_\text{sp}$ always refer to a comoving projected distance and a comoving splashback radius. %In the interest of reproducibility, the code used to produce the analysis presented in this paper is available at \url{github.com/contigiani/kids-splash}.

\section{Data}
\label{sec:data}

This section introduces both the Kilo-Degree Survey \citep[KiDS,][]{deJong2013} and its infrared companion, the VISTA Kilo-degree INfrared Galaxy survey \citep[VIKING,][]{Edge2013}. Their combined photometric catalog and the sample of LRGs extracted from it \citep{2020arXiv200813154V} are the essential building blocks of this paper.

\subsection{KiDS}

KiDs is a multi-band imaging survey in four filters ($ugri$) covering $1350$ deg$^2$. Its fourth data release \citep[DR4, ][]{Kuijken2019DR4} is the basis of this paper and has a footprint of 1006 deg$^2$ split between two regions, one equatorial and the other in the south Galactic cap ($770$ deg$^2$ in total after masking). The $5\sigma$ mean limiting magnitudes in the $ugri$ bands are, respectively, 24.23, 25.12, 25.02, and 23.68. The mean seeing for the $r$-band data, used both as a detection band and for the weak lensing measurements, is 0.7\arcsec. The companion survey VIKING covers the same footprint in five infrared bands, $ZYJHK_s$.

The raw data have been reduced with two separate pipelines, THELI \citep{2005AN....326..432E} for a lensing-optimized reduction of the $r$-band data, and AstroWISE \citep{2013ExA....35...45M}, used to create photometric catalogs of extinction corrected magnitudes. The source catalog for lensing  was produced from the THELI images. Lensfit \citep{2013MNRAS.429.2858M, Conti:2016gav, 2019A&A...624A..92K} was used to extract the galaxy shapes.

\subsection{LRGs}
\label{sec:datalrg}
The LRG sample presented in \cite{2020arXiv200813154V} is based on KiDS DR4. In order to construct the catalogue, the red sequence up to redshift $z=0.8$ was obtained by combining spectroscopic data with the $griZ$ photometric information provided by the two surveys mentioned above. Furthermore, the near-infrared $K_s$ band from VIKING was used to perform a clean separation of stellar objects to lower the stellar contamination of the sample.

The color-magnitude relation that characterizes red galaxies was used to calibrate redshifts to a precision higher than generic photometric-redshift (photo-zs) methods, resulting in redshift errors for each galaxy below $0.02$. For more details on how the total LRG sample is defined and its broad properties, we direct the interested reader to \cite{2020arXiv200813154V}, or \cite{2019MNRAS.487.3715V}, a similar work based on a previous KiDS data release. 

\cite{MCF2021} further analyzed this same catalog and calculated absolute magnitudes for all LRGs using \textsc{LePHARE} \citep{2011ascl.soft08009A} and \textsc{EZGAL} \citep{2012PASP..124..606M}. The first code corrects for the redshift of the rest-frame spectrum in the different passbands  (k-correction), while the second corrects for the passive evolution of the stellar population (e-correction). For this work, we used these (k+e)-corrected luminosities as a tracer of total mass since the two are known to be highly correlated \citep[see, e.g.,][]{2006MNRAS.368..715M, 2015A&A...579A..26V}. Based on this, we then defined two samples with different absolute r-band magnitude cuts, $M_r<-22.8$ and $M_r<-23$, that we refer to as \emph{all} and \emph{high-mass} samples. These are the $10$ and $5$ percentile of the absolute magnitude distribution of the \emph{luminous} sample studied in \cite{MCF2021}, and the two samples contain $5524$ and $2850$ objects each.

\begin{figure}
    \centering
    \includegraphics[width=\columnwidth]{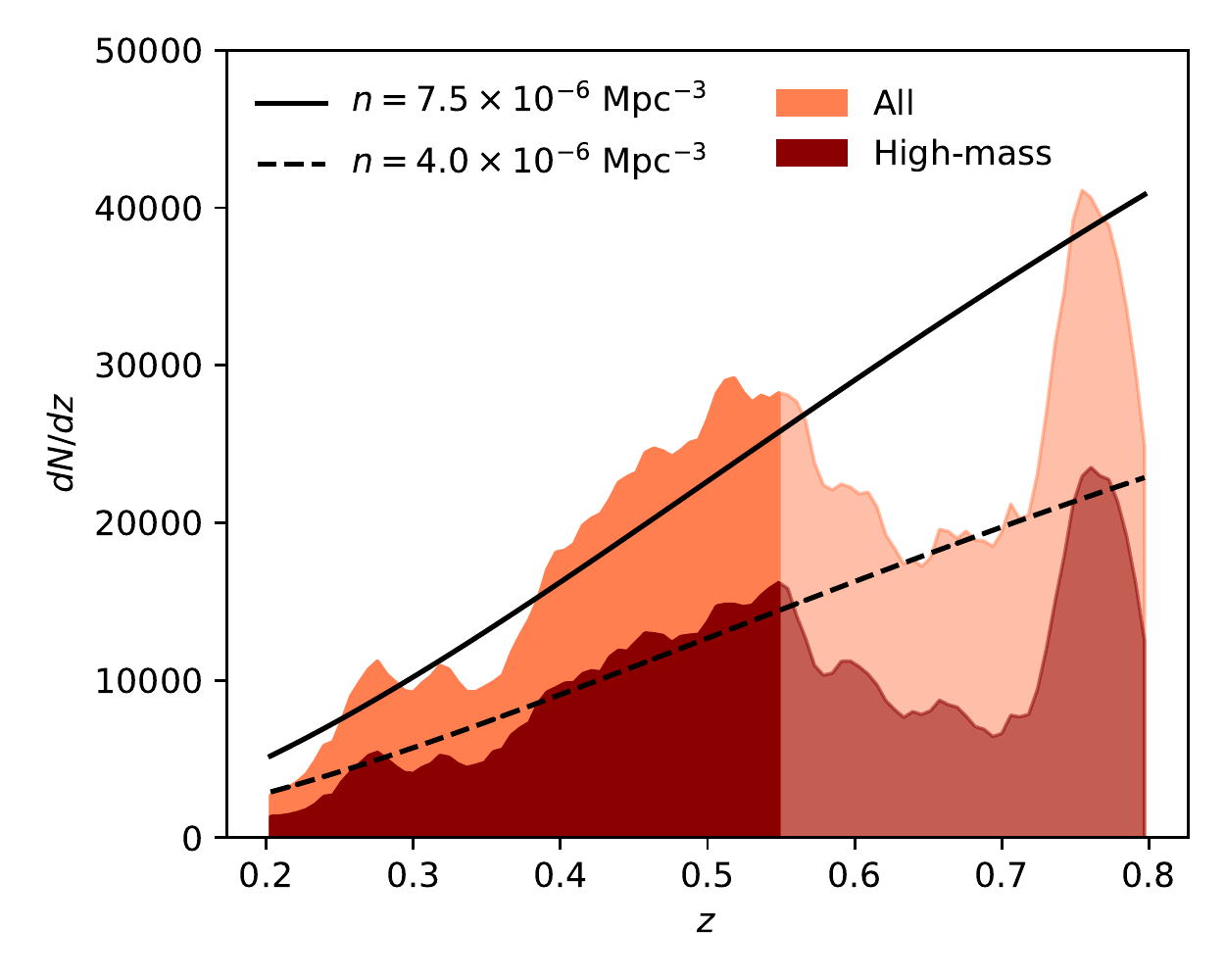}
    \caption{The redshift distributions of the LRG samples studied in this paper. As visible in the figure, the distributions are consistent with the assumption of a constant comoving density up to redshift $z=0.55$, the maximum considered in our main analysis. For higher redshifts, we find that the empirical selection criteria explicitly designed to select for a constant comoving density do not hold. We use the high-redshift tail of our LRG sample (All, $z>0.75$) to investigate the behaviour of our measurements in this regime.}
    \label{fig:redshift}
\end{figure}

Because the (k+e)-correction presented above is designed to correct for observational biases and galaxy evolution, the expected redshift distribution of the LRGs should correspond to a constant comoving density. However, when studying our samples (see Figure~\ref{fig:redshift}), it is clear that this assumption holds only until $z=0.55$. This suggests that the empirical corrections applied to the observed magnitudes are not optimal. It is important to stress that this discrepancy was not recognized before because our particular selection amplifies it: because we consider here the tail of a much larger sample ($N\sim10^5$) with a steep magnitude distribution, a small error in the lower limit induced a large mismatch at the high-luminosity end. To overcome this limitation, we discard all LRGs above $z=0.55$. After fitting the distributions in Figure~\ref{fig:redshift}, we obtained comoving densities $n = 7.5\times 10^{-6}$ Mpc$^{-3}$ and ${n= 4.0\times 10^{-6}~\text{Mpc}^{-3}}$ for the full and the high-mass samples.

\begin{figure}
    \centering
    \includegraphics[width=1\columnwidth]{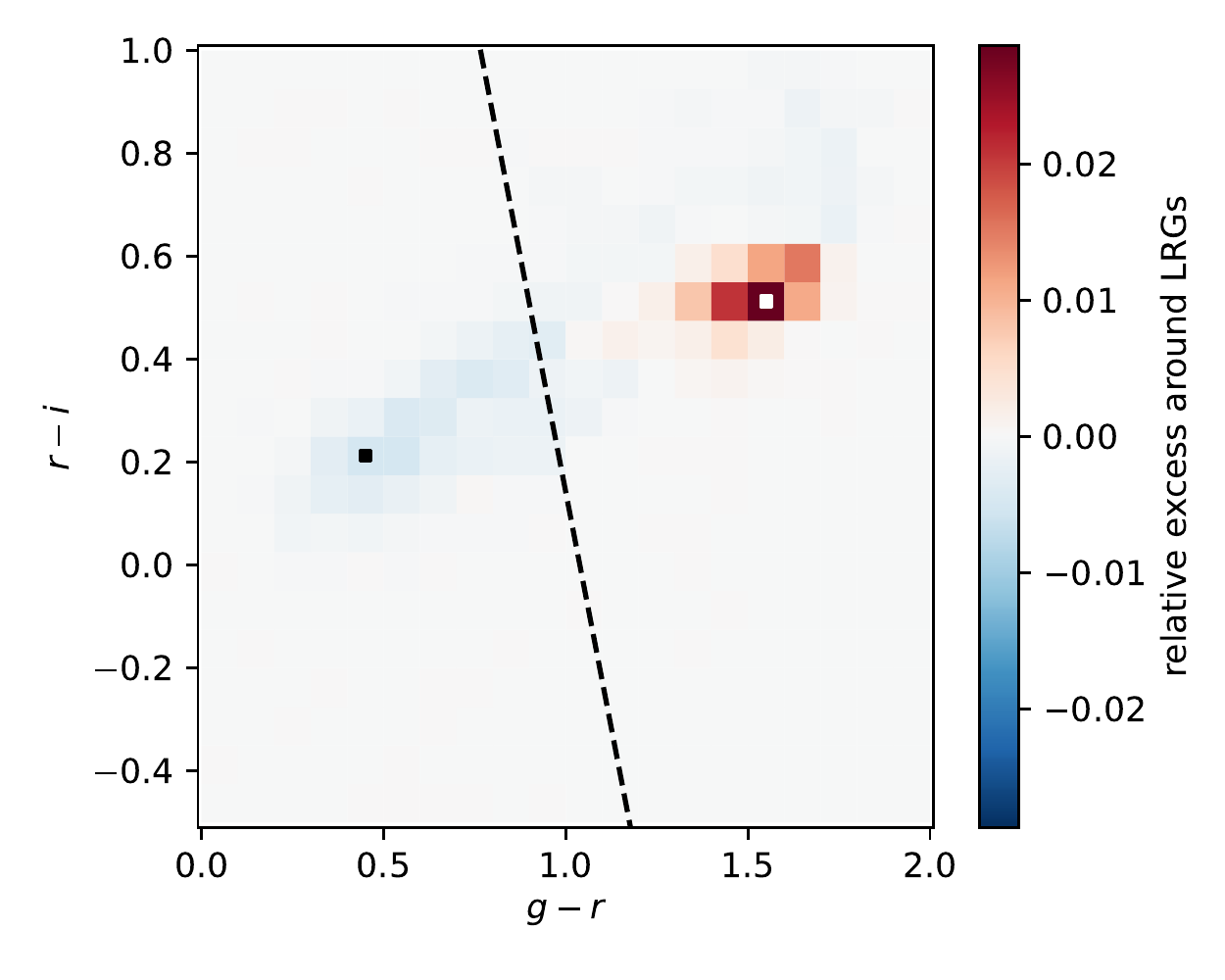}
    \caption{Separating red and blue galaxies. We calculated the distribution of KiDS galaxies in the ($g$-$r$)-($r$-$i$) color plane for objects around random points in the sky and around LRGs in the \emph{high-mass} sample between redshifts $0.3$ and $0.35$ ($R<1$ Mpc). This histogram represents the difference between the two distributions as a fraction of the entire KiDS population. The black and white squares mark the pixels with the lowest and highest value. An overdensity of red objects and an underdensity of blue objects is apparent, and the line separating the two locations is used to split the full KiDS sample into two populations.}
    \label{fig:color_split}
\end{figure}

\begin{figure*}
    \centering
    \includegraphics[width=1\columnwidth]{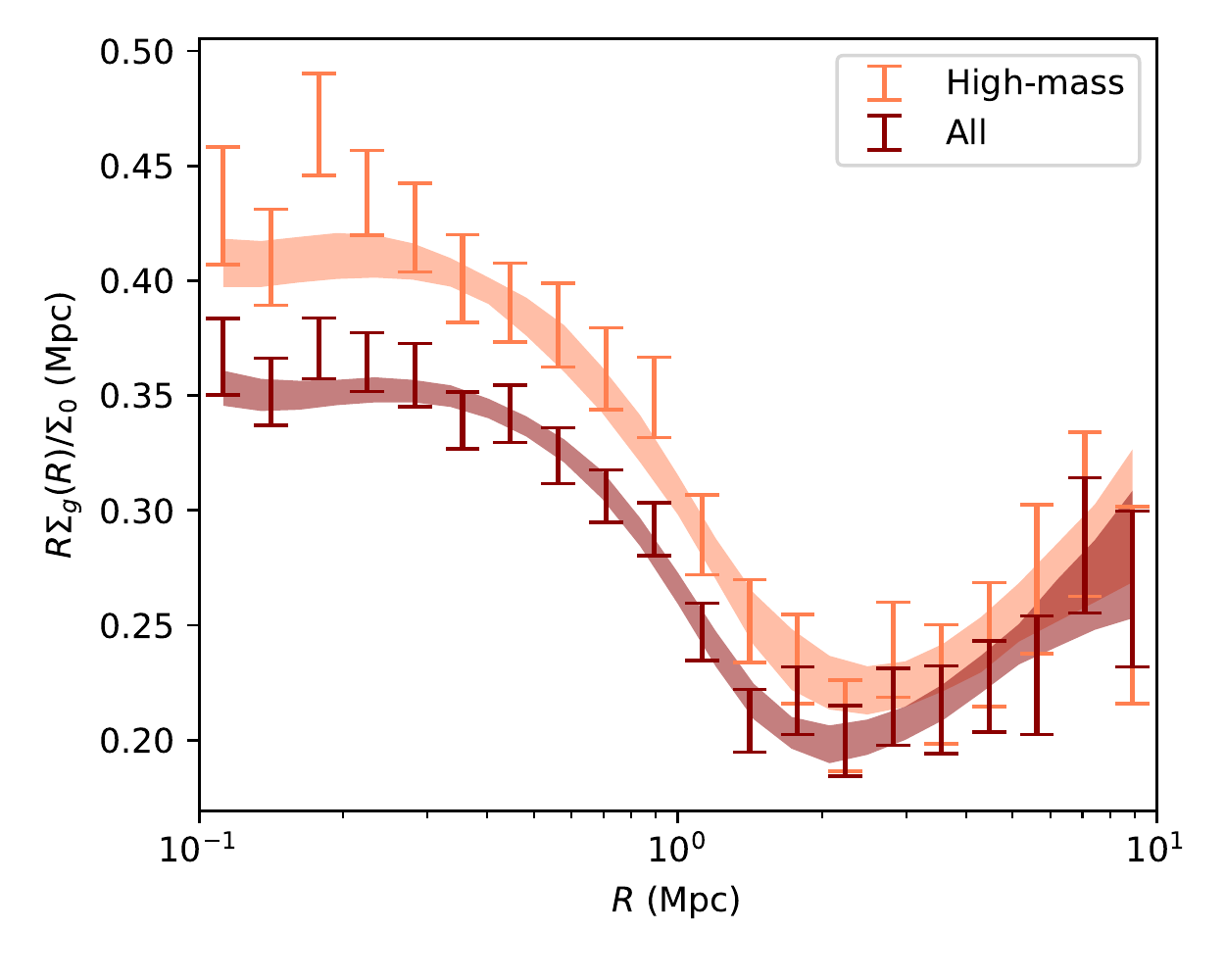}
    \includegraphics[width=1\columnwidth]{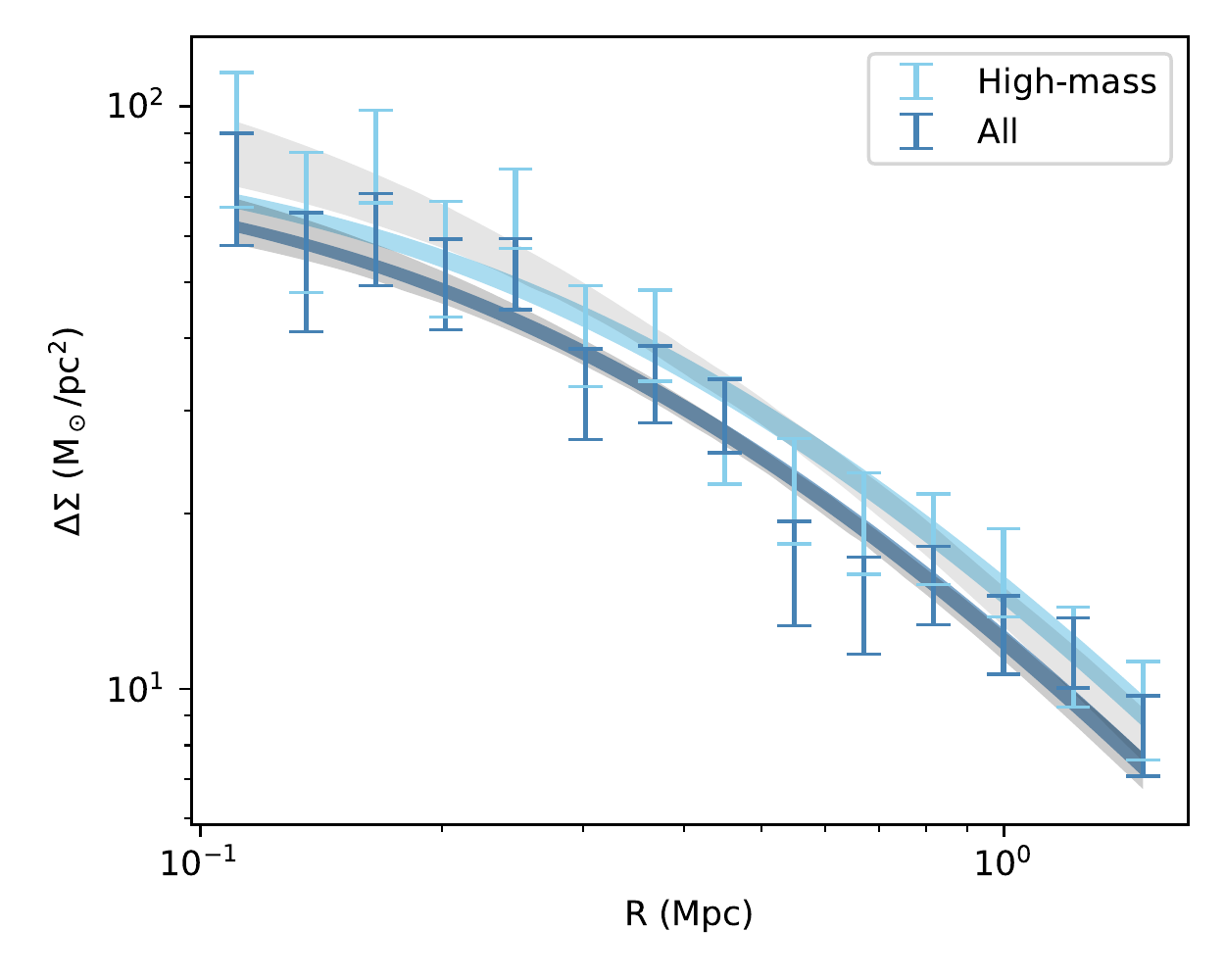}
    \caption{The signals studied in this paper. We measure the number density of KiDs red galaxies (left panel) and the lensing signal (right panel) around the LRGs in our sample (\emph{all}) and its high-luminosity subsample (\emph{high-mass}). Both measurements are based on the KiDS photometric catalog. The steep drop around ${1~\text{Mpc}}$ visible in the left panel is the splashback feature, and it is connected to the total mass of the LRG haloes. Similarly, the amplitude of the lensing signal on the right is also a measure of the same mass. In addition to the data and the $1\sigma$ error bars, we also display the $68$ percent contours of two profile fits performed to extract the mass measurements. The fit on the right is performed either by varying only the amplitude of the signal (thinner contours) or by varying its amplitude and concentration (wider contours). See text for more details. Section~\ref{sec:data} presents the data and the two samples, Section~\ref{sec:profiles} discusses how the profiles are measured, and Section~\ref{sec:fit} discusses the fitting procedure.}
    \label{fig:measurement}
\end{figure*}

\section{Profiles}
\label{sec:profiles}

In this section, we discuss how we used our data sets to produce two stacked signals measured around the LRGs: the galaxy profile, capturing the distribution of fainter red galaxies, and the weak lensing profile, a measure of the projected mass distribution extracted from the distorted shapes of background galaxies. We present these two profiles and the $68$ percent contours of two separate parametric fits in Figure~\ref{fig:measurement}. The details of the fitting procedure are explained in Section~\ref{sec:fit}.

\subsection{Galaxy profile}
\label{sec:galaxyanalysis}

We expect bright LRGs to be surrounded by fainter satellites, i.e., we expect them to be the central galaxies of galaxy groups or clusters. To obtain the projected number density profile of the surrounding KiDS galaxies, we split the LRG samples in $7$ redshift bins of size $\delta_z = 0.05$ in the range $z\in[0.2, 0.55]$. We then defined a corresponding KiDS galaxy catalog for each redshift bin, obtained the background-subtracted distribution of these galaxies around the LRGs, and finally stacked these distributions using the weights $w_i$ defined below. 

We did not select the KiDS galaxies by redshift due to their large uncertainty. Instead, for each redshift bin, we used the entire KiDS catalogs and only applied two redshift-dependent selections: one in magnitude and one in color space. The reason behind the first selection is simple: compared to a flat signal-to-noise ratio (SNR) threshold, a redshift-dependent magnitude limit does not mix populations with different intrinsic magnitudes as a function of redshift \citep[as suggested by][]{2016ApJ...825...39M}. On the other hand, the color cut has a more physical explanation. Red satellites are the most abundant population in galaxy clusters and, due to their repeated orbits inside the host cluster, they are known to better trace dynamical features such as splashback  \citep[see, e.g.,][]{2017ApJ...841...18B}. Combining these two criteria also has the effect of selecting a similar population even in the absence of k-corrected magnitudes. 

For the highest redshift considered here, $z_\text{max}$, we limited ourselves to observed magnitudes $m_r<23$, equivalent to a $10$ SNR cut. We then extrapolated this limit to other redshift bins by imposing
\begin{equation}
    \label{eq:magcut}
    m_r < 23 - 5\log \left(
    \frac{d_L(z_\text{max})}{d_L(z_i)} \right),
\end{equation}
where $z_i$ is the upper edge of the redshift bin considered, and $d_L(z)$ is the luminosity distance as a function of redshift. Afterward, we divided the galaxy catalogs into two-color populations by following the method of \cite{2020arXiv200811663A}. Compared to random points in the sky, the color distribution of KiDS galaxies around LRGs contains two features: an overdensity of ''red`` objects and a deficit of ''blue`` objects. Based on the red-sequence calibration of \cite{2020arXiv200813154V} and the location of the  $4000$ \AA~break, we identified the ${(g-r)-(r-i)}$ plane as the most optimal color space to separate these two populations at redshifts $z\leq 0.55$. We also noted that the ${(i-Z)-(r-i)}$ plane would be better suited for higher redshifts. From the distribution in the color-color plane, the two classes can then be separated by the line perpendicular to the segment connecting these two loci and passing through their midpoint. Figure~\ref{fig:color_split} provides an example of this procedure. We point out that a more sophisticated selection could be used since the structure in color space suggests the existence of a compact red cloud. For the purposes of this work, however, we do not find this to be necessary. 

We used \textsc{treecorr} \citep{2004MNRAS.352..338J, 2015ascl.soft08007J} to extract the correlation functions from the red galaxy catalogs defined above
\begin{equation}
    \xi_i = \frac{DD_i}{DR_i} - 1,
\end{equation}
where $DD$ and $DR$ are the numbers of LRG-galaxy pairs calculated using the KiDS catalogs or the random catalogs, respectively. These randoms are composed of points uniformly distributed in the KiDS footprint. The error covariance matrices of these measurements were obtained by dividing the survey area into $50$ equal-areal jackknife regions. Because the signal is statistics limited, the off-diagonal terms of this matrix are found to be negligible. To further support this statement, we point out that due to the low number density of the sample (see Figure~\ref{fig:redshift}), the clusters do not overlap in real space.

Formally, the correlation function written above is related to the surface overdensity of galaxies:
\begin{equation}
    \Sigma_{i}(R)  = \xi_i(R) \Sigma_{0, i},
\end{equation}
where $\Sigma_{0, i}$ is the average surface density of KiDS galaxies in the $i$-th redshift bin. However, since we are interested in the shape of the profile and not its amplitude, we did not take this parameter into account when stacking the correlation functions $\xi_i$. 
The signal considered in this paper is a weighted sum of the individual correlation functions. Formally:
\begin{equation}
    \frac{\Sigma_\text{g}(R)}{\Sigma_0} = \frac{\sum_i w_i(R) ~\xi_i(R)}{\sum_i w_i(R)},
\end{equation}
where $\Sigma_0$ is a constant needed to transform the dimensionless correlation function into the projected mass density. Because we decided to fit the combination $\Sigma_\text{g}(R)/\Sigma_0$ directly, the value of this constant is unimportant. To optimize the stacked signal, we used as weights $w_i$ the inverse variance of our measurement. This corresponds to an SNR weighted average, where the SNR is, in our case, dominated by the statistical error of the DD counts.

The left side of Figure~\ref{fig:measurement} presents our measurement of the galaxy profile around the LRGs. As expected, the high-mass subsample has a higher amplitude compared to the entire sample. 

\subsection{Weak lensing profile}
\label{sec:lensinganalysis}

The shapes of background sources are deformed, i.e., lensed, by the presence of matter along the line of sight. In the weak lensing regime, this results in the observed ellipticity $\bm{\epsilon}$ of a galaxy being a combination of its intrinsic ellipticity and a lensing shear. If we assume that the intrinsic shapes of galaxies are randomly oriented, the coherent shear in a region of the sky can therefore be computed as the mean of the ellipticity distribution. %This mean is known as the lensing signal. 

Consider a circularly symmetric matter distribution acting as a lens. In this case, the shear only contains a tangential component, i.e., the shapes of background galaxies are deformed only in the direction parallel and perpendicular to the line in the sky connecting the source to the center of the lens. Because of this, we can define the lensing signal in an annulus of radius $R$ as the average value of the tangential components of the ellipticities $\epsilon^{(t)}$. The next few paragraphs provide the details of the exact procedure we followed to measure this lensing signal around the LRGs in our samples. For this second measurement, we used the weak lensing KiDS source catalog extending up to redshift $z=1.2$ \citep[see also,][]{2015MNRAS.452.3529V, Dvornik_2017}. 

Based on the lensfit weights $w_s$ associated with each source, we defined \emph{lensing} weights for every lens-source combination,
\begin{equation}
    \label{eq:lensingeff}
    w_\text{l,s} = w_\text{s} \left(\Sigma_{\text{crit, l}}^{-1}\right)^{2}, 
\end{equation}
where the two indices $\text{l}$ and $\text{s}$ are used to indicate multiple lens-source pairs. The second factor in the product above represents a lensing efficiency contribution and, in our formalism, this quantity  does not depend on the source. It is calculated instead as an average over the entire source redshift distribution $n(z_\text{s})$: 
\begin{equation}
    \label{eq:lensingeff2}
    \Sigma_\text{crit, l}^{-1} = \frac{4\pi G}{c^2} \frac{d_\text{A}(z_\text{l})}{(1+z_\text{l})^2} \int_{z_\text{l}+\delta}^{\infty} dz_\text{s} \; \frac{d_\text{A}(z_\text{l}, z_\text{s})}{d_\text{A}(0, z_\text{s})} n(z_\text{s}),
\end{equation}
where $d_\text{A}(z_1, z_2)$ is the angular diameter distance between the redshifts $z_1$ and $z_2$ in the chosen cosmology. Sources that belong to the correlated structure surrounding the lens might scatter behind it due to the uncertainty of the photometric redshifts. The gap between the lens plane and the source plane in the expression above ($\delta=0.2$) ensures that our signal is not diluted by this effect \citep[see appendix A4 of][]{Dvornik_2017}. 
%The factor (1+$z_\text{l}$) in this expression is introduced because we are working in comoving coordinates. 
Once all of these ingredients are computed, an estimate of the measured lensing signal is given by:
\begin{equation}
\label{eq:deltaS}
   \Delta \Sigma (R) = 
   \frac{
    \sum_\text{l,s}  \epsilon^\text{(t)}_{\text{l,s}} w_\text{l,s} \Sigma_{\text{crit, l}}
                        }{ 
    \sum_\text{l,s} w_\text{l,s}
                        } 
    \frac{1}{1+m},
\end{equation}
where the sums are calculated over every source-lens pair, and $m$ is a residual multiplicative bias of order $0.014$ calibrated using image simulations \citep{Conti:2016gav, 2019A&A...624A..92K}. This signal is connected to the mass surface density $\Sigma_\text{m}(R)$ and its average value within that radius, $\overline{\Sigma}_\text{m}(<R)$.
\begin{equation}
\label{eq:esd}
    \Delta \Sigma (R) = \overline{\Sigma}_\text{m}(<R) - \Sigma_\text{m} (R).
\end{equation}

The covariance matrix of this average lensing signal was extracted through bootstrapping, i.e., by resampling $10^5$ times the $1006$ $1\times1$ deg$^2$ KiDS tiles used in the analysis. This signal, like the galaxy profile before, is also statistics limited. Therefore we have not included the negligible off-diagonal terms of the covariance matrix in our analysis. 

Finally, we note that we have thoroughly tested the consistency of our lensing measurement. We computed the expression in Equation~\eqref{eq:deltaS} using the cross-component $\epsilon^{(\times)}$ instead of the tangential $\epsilon^\text{(t)}$ and verified that its value was consistent with zero. Similarly, we also confirmed that the measurement was not affected by additive bias by measuring the lensing signal evaluated around random points.

\section{Three ways to measure cluster masses}
\label{sec:fit}

This section presents three independent measures of the total mass contained in the LRG haloes. We refer to these estimates as splashback (or dynamical) mass, lensing mass and abundance mass. The first two are extracted by fitting parametric profiles to the two signals presented in the previous section (Figure~\ref{fig:measurement}), and the third is based on a simple abundance matching argument. Fitting the galaxy profile allows us to constrain the splashback feature and provides a dynamical mass, while fitting the amplitude of the lensing signal provides a lensing mass.

\begin{table}
\begin{center}
\begin{tabular}{c|c|c}
    \hline
    Parameter & Prior \\ \hline
    $\alpha$ & $\mathcal{N}(0.2, 2)$ \\
     $g$ & $\mathcal{N}(4, 0.2)$  \\  
     $\beta$ & $\mathcal{N}(6, 0.2)$  \\  
     $r_\text{t}/(1~\text{Mpc})$ & $\mathcal{N}(1, 4)$  \\  
     $s_\text{e}$ & $[0.1, 2]$   \\ \hline
\end{tabular}
\end{center}
\caption{The priors used in the fitting procedure of Section~\ref{sec:fit}. When fitting the data in the left panel of Figure~\ref{fig:measurement}, we employ the model in Equation~\eqref{eq:DK14} with the priors presented above. For some parameters, we impose flat priors in a range, e.g. $[a, b]$, while for others we impose a Gaussian prior $\mathcal{N}(m, \sigma)$ with mean $m$ and standard deviation $\sigma$. We do not restrict the prior range of the two degenerate parameters $\bar{\rho}$ and $r_0$.}
\label{tab:priors}
\end{table}

\begin{figure*}
    \centering
    \includegraphics[width=1\columnwidth]{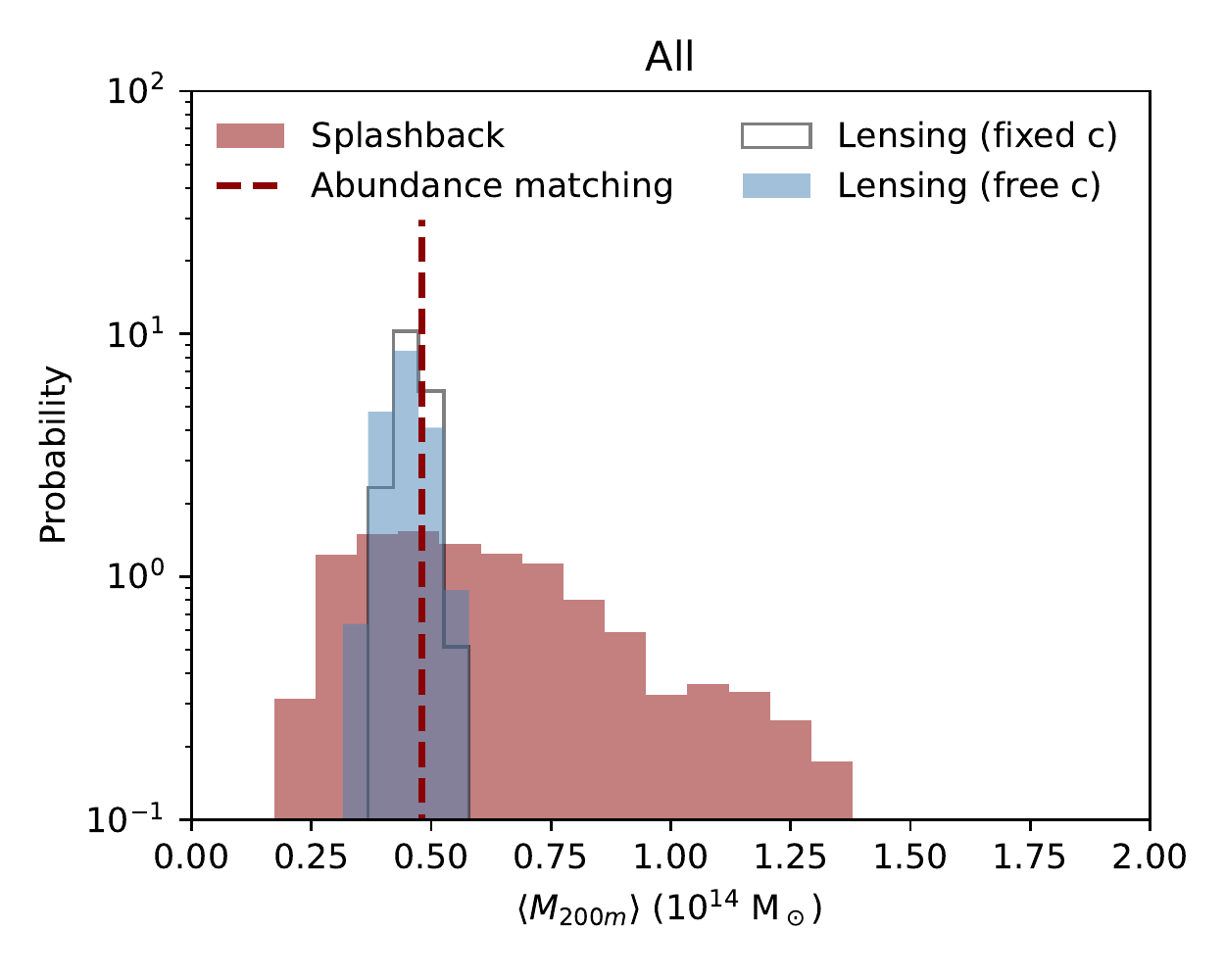}
    \includegraphics[width=1\columnwidth]{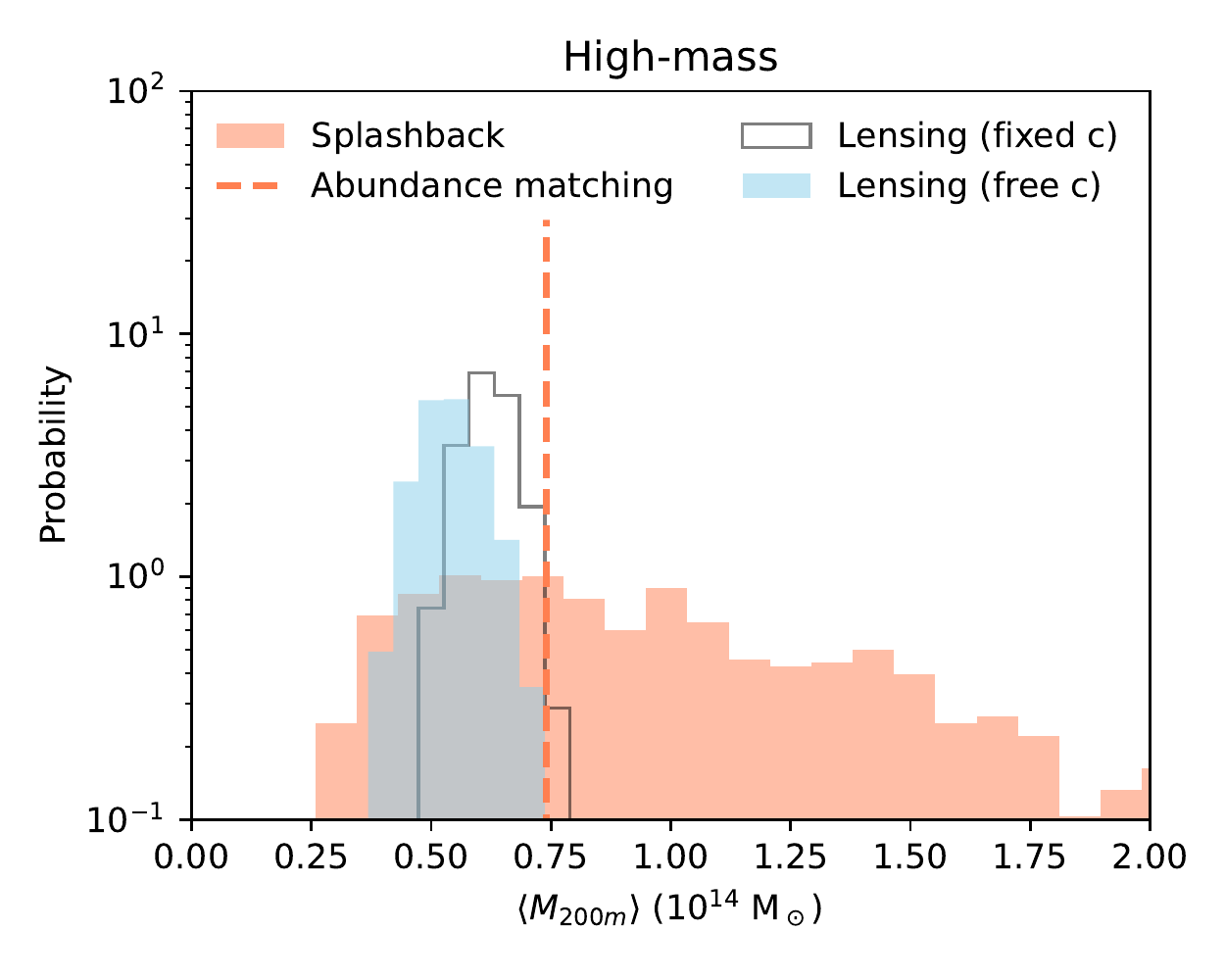}
    \caption{Comparison of the mass measurements performed in this paper. Using three different techniques, we measured the mass of the haloes hosting our LRG sample (\emph{all}) and a high-luminosity subsample (\emph{high-mass}). The remarkable consistency between the three methods for both samples is a testament to the robustness of our LRG selection and the prospect of measuring halo masses from the splashback feature. Table~\ref{tab:table_masses} reports the same results in textual form. See Section~\ref{sec:discussion} for more details about this comparison.} 
    \label{fig:mass}
\end{figure*}

\subsection{Splashback mass}
\label{sec:fitmass}

Thanks to the splashback feature, it is possible to estimate the total halo mass by fitting the galaxy distribution with a flexible enough model. The essential feature that such three-dimensional profile, $\rho(r)$, must capture is a sudden drop in density around $r_\text{200m}$. Its most important parameter is the point of steepest slope, also known as the splashback radius $r_\text{sp}$. Equivalently, this location can be defined as the radius where the function $d \log \rho/d \log r$ reaches its minimum. 

In general, the average projected correlation function can be written in terms of the average three-dimensional mass density profile as:
\begin{equation}
\label{eq:S}
\frac{\Sigma_\text{g} (R)}{\Sigma_0} = \frac{2}{\Sigma_0}\int_0^{\infty} d\Delta \, \rho\left(\sqrt{\Delta^2 + R^2}\right),
\end{equation}
In practice, we evaluated this integral in the range [$0$, $40$] Mpc and confirmed that our results are not sensitive to the exact value of the upper integration limit. 

The specific density profile that we have used is based on \cite{2014ApJ...789....1D}, and it has the following form:
\begin{align}
\label{eq:DK14}
    \rho(r) &= 
    \rho_{\text{Ein}}(r) f_{\text{trans}} (r) + \rho_{\text{out}} (r), \\
    \rho_{\text{Ein}}(r) &= \rho_{\text{s}} \exp \left( -\frac{2}{\alpha}\left[\left( \frac{r}{r_{\text{s}}} \right)^{\alpha} - 1  \right] \right), \\
    f_{\text{trans}} (r) &=  \left[ 1+ \left(\frac{r}{r_\text{t}}\right)^{\beta}\right]^{-g/\beta}, \\
    \rho_{\text{out}} &= \bar{\rho} \left( \frac{r}{r_0} \right)^{-s_\text{e}}.
\end{align}

These expressions define a profile with two components: an inner halo and an infalling region. 
The term $\rho_\text{Ein}(r) f_\text{trans}(r)$ represents the collapsed halo through a truncated Einasto profile with shape parameter $\alpha$ and amplitude $\rho_s$ \citep{Einasto1965}. 
The parameters $g, \beta$ in the transition function determine the maximum steepness of the sharp drop between the two regions, and $r_\text{t}$ determines its approximate location. Finally, the term $\rho_\text{out}(r)$ describes a power-law mass distribution with slope $s_\text{e}$ and amplitude $\bar{\rho}$, parametrizing the outer region dominated by infalling material. For more information about the role of each parameter and its interpretation, we refer the reader to \cite{2014ApJ...789....1D}, and previous measurements presented in the introduction \citep[see, e.g.,][for more details about the role of the truncation radius $r_\text{t}$]{2019MNRAS.485..408C}. 

This profile is commonly used to parameterize mass profiles but is used in this section to fit a galaxy number density profile. When performing this second type of fit, the amplitudes $\rho_\mathrm{s}$ and $\bar{\rho}$ are dimensionless and, together with the flexible shape of the profile, completely capture the connection between the galaxy and matter density fields. Similarly to $\Sigma_0$, the value of these constants is not the focus of this paper. 
 
To extract the location of the splashback radius for our two LRG samples, we fitted this model profile to the correlation function data using the ensemble sampler \textsc{emcee} \citep{Foreman-Mackey2013}. The priors imposed on the various parameters are presented in Table~\ref{tab:priors}, and we highlight in particular that the range for $\alpha$ is a generous scatter around the expectation from numerical simulations \citep{Gao2008}. The best-fitting profiles extracted from this procedure are shown in Figure~\ref{fig:measurement}.

In clusters, the location of the central galaxy might not correspond to the barycenter of the satellite distribution. While this discrepancy is usually accounted for in the modeling of the projected distribution in Equation~\eqref{eq:S}, we chose not to consider this effect in our primary analysis. This is justified by the fact that the miscentering term affects the profile within $R\sim0.1$ Mpc, while we are interested in the measurement around $R\sim 1$ Mpc \citep{2021arXiv210505914S}, and the data do not require a more flexible model to provide a good fit. 

Finally, to transform the $r_\text{sp}$ measurements into a value for $M_\text{200m}$, we used the relations from \citet{2020ApJS..251...17D}, evaluated at our median redshift of $\bar{z}=0.44$. In this transformation, we employed the suggested theoretical definition of splashback, based on the $75$th percentile of the dark matter apocenter distribution.  In the same paper, this definition of splashback based on particle dynamics has been found to accurately match the definition based on the  minimum of $\log \rho / \log r$ used in this work. For more details about the relationship between these two definitions, we refer the reader to section 3.1 of \cite{2021MNRAS.tmp.1404C}. 

Because the splashback radius depends on accretion rate, we used the median value of this quantity as a function of mass as a proxy for the effective accretion rate of our stacked sample. We note in particular that the additional scatter introduced by the accretion rate and redshift distributions is expected to be subdominant given the large number of LRGs we have considered.

\subsection{Lensing mass}

To extract masses from the lensing signal, we performed a fit using an NFW profile \citep{Navarro1996, Navarro1997}:
\begin{equation}
\label{eq:NFW}
    \rho(r) = 
     \frac{1}{4 \pi F(c_\text{200m})} 
    \frac{M_\text{200m}}{r(r+ r_\text{200m}/c_\text{200m})^2},
\end{equation}
where $M_\text{200m}$ and $r_\text{200m}$ are related by Equation \eqref{eq:200m}, $c_\text{200m}$ is the halo concentration, and the function appearing in the first term is defined as:
\begin{equation}
\label{eq:f}
    F(c) =\ln(1+c)-c/(1+c).
\end{equation}
From this three-dimensional profile, the lensing signal can be derived by replacing $\Sigma_\text{g}/\Sigma_0$ with $\Sigma_\text{m}$ in the projections Equations~\eqref{eq:esd} and \eqref{eq:S}.

We point out that we did not use the complex model of Equation~\eqref{eq:DK14} for the lensing measurement. This is because, the differences between the Einasto profile used there and the NFW profile presented above are not expected to induce systematic biases at the precision of our measurements \citep[see, e.g.,][]{2016JCAP...01..042S}. Although extra complexity might not be warranted, particular care should still be taken when measuring profiles at large scales, where the difference between the more flexible profile and a traditional NFW profile is more pronounced. Consequently, we reduce any bias in our measurement by fitting only projected distances $R<1.5$ Mpc, where the upper limit is decided based on the $r_\text{sp}$ inferred by our galaxy distribution measurement.

Since the mass and concentration of a halo sample are related, several mass-concentration relations calibrated against numerical simulations are available in the literature. 
For the measurement presented in this section, we used the mass-concentration relation of \cite{2013ApJ...766...32B}. However, because this relation is calibrated with numerical simulations based on a different cosmology, we also fit the lensing signal while keeping the concentration as a free parameter. This consistency check is particularly important because halo profiles are not perfectly self-similar \citep{2015ApJ...799..108D} and moving between different cosmologies or halo mass definitions might require additional calibration. 
We perform the fit to the profiles in the right panel of Figure~\ref{fig:measurement} using the median redshift of our samples, $\bar{z}=0.44$. We find that statistical errors dominate the uncertainties, and we do not measure any systematic effect due to the assumed mass-concentration relation.

\subsection{Abundance mass}

In addition to the two mass measurements extracted from the galaxy and lensing profiles, we also calculated masses using an abundance matching argument. 

The comoving density of haloes of a given mass is a function of cosmology \citep{1974ApJ...187..425P}. Since we expect a tight relationship between the mass of a halo and the luminosity of the associated galaxy, any lower limit in the first can be converted into a lower limit in the second. Therefore, our measurement of the comoving density in Figure~\ref{fig:redshift} can be converted into a mass measurement. We note, in particular, that this step assumes that \citet{2020arXiv200813154V} built a complete sample of LRGs with no contamination and that the luminosity estimates obtained in \citet{MCF2021} are accurate, at least in ranking.

We used the mass function of \cite{2008ApJ...688..709T} at the median redshift $\bar{z}=0.44$ to convert our fixed comoving densities into lower limits on the halo mass $M_\text{200m}$. To complete the process, we then extracted the mean mass of the sample using the same 
mass function. 

The relation between halo mass and galaxy luminosity is not perfect, however, since the galaxy luminosity function is shaped by active galactic nuclei activity and baryonic feedback. These processes induce an increased scatter in the stellar mass to halo mass relation \citep{2014MNRAS.445..175G}, which we have not accounted for. This effect, combined with the uncertainties in the LRG selection and luminosity fitting, are the main sources of error for our abundance matching mass. Since we have not performed these steps in this work, however, we decided not to produce an uncertainty for this measurement and report it here without an error bar.

\begin{figure}
    \centering
    \includegraphics[width=1\columnwidth]{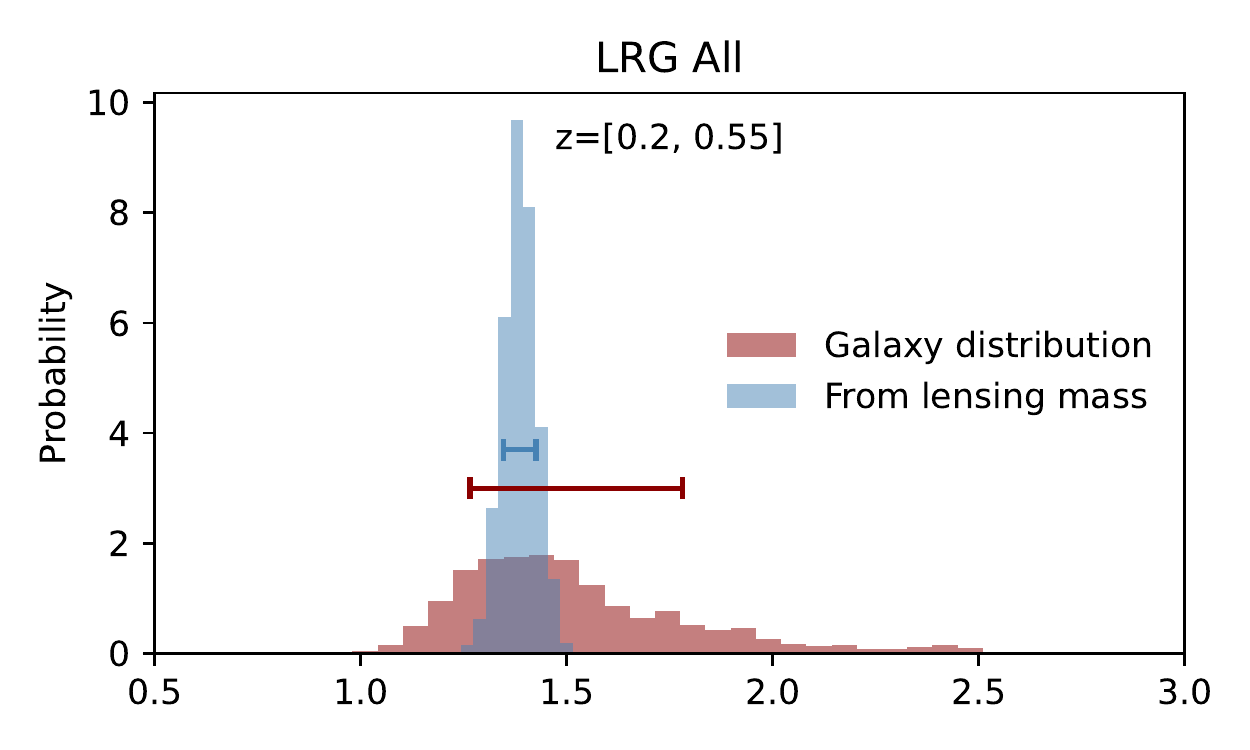}
    \includegraphics[width=1\columnwidth]{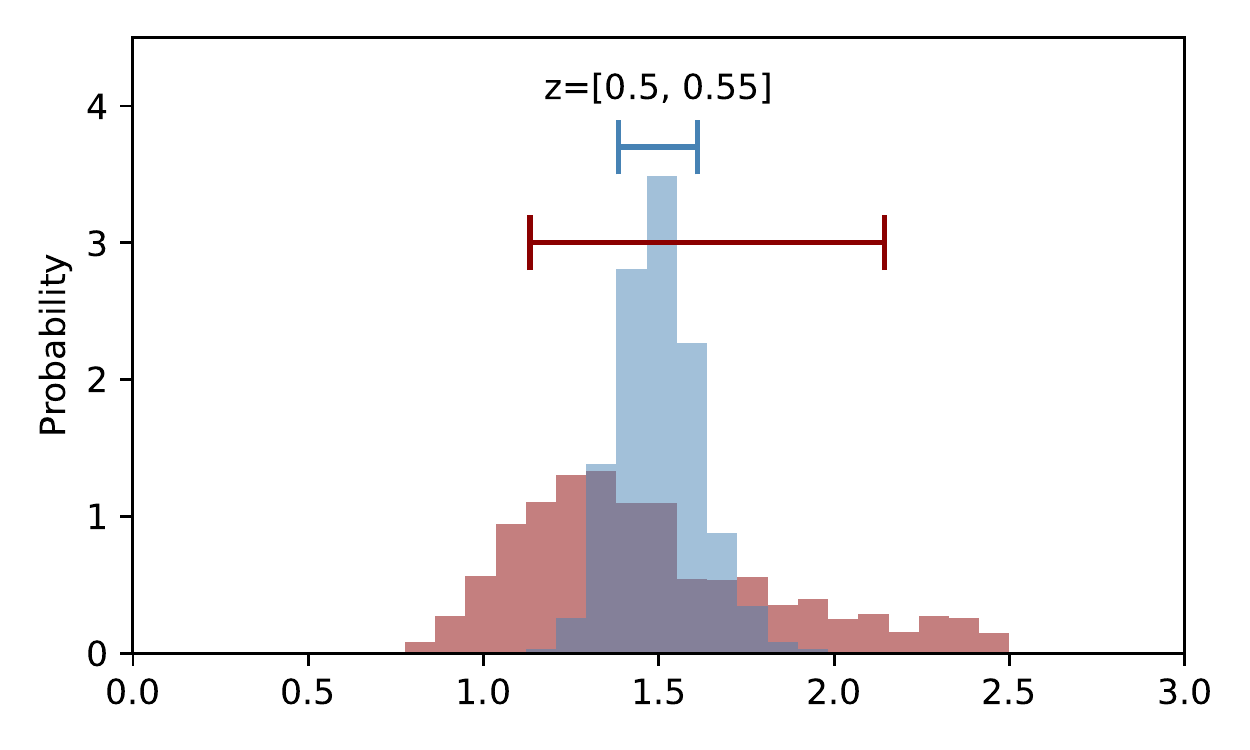}
    \includegraphics[width=1\columnwidth]{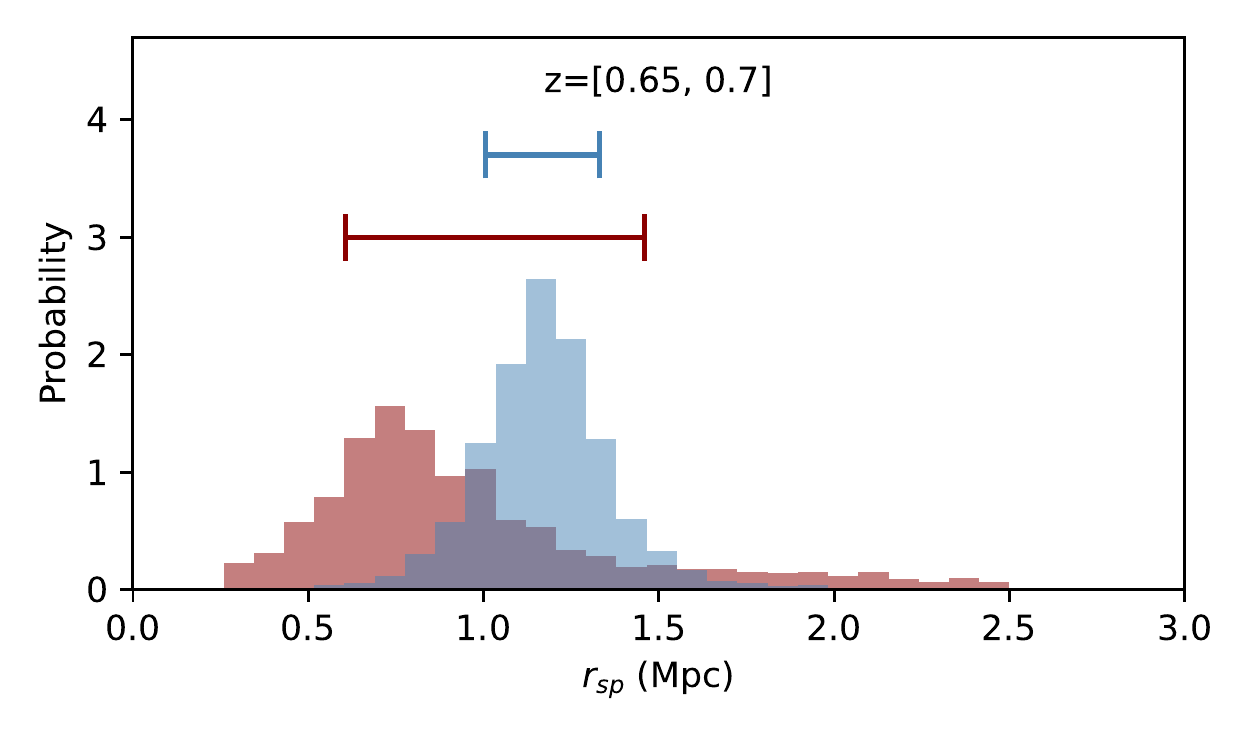}
    \caption{Galaxy distribution measurements scale better with redshift compared to lensing measurements. The three panels show the posterior distribution of $r_\text{sp}$ obtained with the two techniques discussed in this paper for three different redshift ranges. The colored error bars indicate the $68$ percentile interval of each distribution. From top to bottom the ratio between the intervals for the two techniques increases significantly: $[0.15, 0.25, 0.4]$, proving the presence of a different redshift dependence that benefits the galaxy distribution measurement. Note that this figure uses $r_\text{sp}$ as a comparison variable instead of the mass used in Figure~\ref{fig:mass}. This choice is due to the smaller error bars for this parameter. See the final paragraph of
    Section~\ref{sec:fitmass} for more details about how the one-to-one transformation between these two variables was obtained.}
    \label{fig:hz}
\end{figure}

\section{Discussion}
\label{sec:discussion}

In this section, we compare and validate the measurements presented in the previous one. As an example of the power granted by multiple cluster mass measurements from the same survey, we also present an interpretation of these measurements in the context of modified theories of gravity. 

In Figure~\ref{fig:mass} and Table~\ref{tab:table_masses}, we present the results of our two main mass measurements combined with the abundance-matching estimate introduced in the previous subsection. All measurements are in agreement, providing evidence that there is no significant correlation between the selection criteria of our LRG sample and the measurements performed here. The inferred average splashback masses of our LRG samples have an uncertainty of around $50$ percent. 

The first striking feature is the varying degree of precision among the different measurements. The lensing result is the most precise, even when the concentration parameter is allowed to vary. In particular, the fact that the inferred profiles do not exhaust the freedom allowed by error bars in the right-hand panel of Figure~\ref{fig:measurement}
implies that our NFW model prior is responsible for the strength of our measurement and that a more flexible model will result in larger mass uncertainties. On the other hand, with splashback, we can produce a dynamical mass measurement without any knowledge of the shape of the average profile and, more importantly, without having to capture the exact nature of the measured scatter. 

There is also a second, more important, difference between the two measurements that we want to highlight here. The SNR of the splashback mass is dominated by high-redshift LRGs since $\text{SNR}\sim \sqrt{N_\text{LRG}}$. While the ability to capture intrinsically fainter objects at low redshift might affect this scaling, we point out that the redshift-dependent magnitude cut introduced in Equation~\eqref{eq:magcut} explicitly prevents this. In contrast, the lensing weights in Equations~\eqref{eq:lensingeff} imply that the more numerous high-redshift objects do not dominate the lensing signal. This is due to a combination of the lower number of background sources available, the lower lensfit weights associated with fainter sources, and the geometrical term in Equation~\eqref{eq:lensingeff2}. 

This point is explored quantitatively in Figure~\ref{fig:hz}, where we compare the two techniques for different redshift bins. The top panel is a projection of the left-hand panel of Figure~\ref{fig:mass} in terms of $r_\text{sp}$, while the other two are new results. These new measurements at higher redshift are obtained using the same methods presented in Section~\ref{sec:profiles}. To be precise: for the galaxy distribution, we impose a $10$ SNR cut for the KiDS galaxies and a subsequent color selection in the ${(i-Z)-(r-i)}$ plane; while for the lensing signal, we use the same source selection presented before. As visible in the Figure, both measurements degrade for higher redshifts, but the two scale differently. If we consider the size of the $68$ percentile intervals for the two measurements,  at $z=[0.2, 0.5]$ we obtain a ratio between the two of $1:7$, while at $z=[0.65, 0.7]$ we obtain a ratio of $1:2.5$, significantly better. As discussed in a future section, this different scaling has important implications for future photometric missions.

As a final note on our main results, we point out that the difference between the masses of the two samples (\emph{all} and \emph{high-mass}) is $2\sigma$ for the lensing measurement, but it is not even marginally significant for the splashback values (due to the large error bars). As already shown in \cite{2019MNRAS.485..408C}, splashback measurements are heavily weighted towards most massive objects. To produce a non-mass weighted measure of the splashback feature, it is necessary to rescale the individual profiles with a proxy of the halo mass. However, because the study of $r_\text{sp}$ as a function of mass is not the main focus of this work, we leave this line of study open for future research.

\begin{table}
    \hspace{-0.7cm}
    \begin{tabular}{l|c|c|c|c}
    \hline
       Technique  & \multicolumn{2}{c}{$M_\text{200m}$ ($10^{14}$ M$_\odot$)}  & \multicolumn{2}{c}{$r_\text{sp}$ (Mpc)} \\
       & All & High-mass & All & High-mass \\
       \hline
%Splashback & $0.57^{+0.73}_{-0.22}$ & $0.77^{+0.64}_{-0.30}$ & $1.48\pm 0.28$ & $1.6\pm 0.25$ \\ 
Splashback & $0.57^{+0.36}_{-0.21}$ & $0.9^{+0.85}_{-0.38}$ & $1.48\pm 0.2$ & $1.68\pm 0.28$ \\
         Lensing (fixed c) & $0.46\pm 0.03$ & $0.62\pm 0.05$ & $1.40\pm 0.01$ & $1.52\pm 0.02$  \\ 
         \hline
         Lensing (free c) & $0.44\pm 0.05$ & $0.54\pm 0.07$ & $1.39\pm 0.03$ & $1.6\pm 0.04$  \\ 
         Abundance & $0.48$ & $0.74$ & $1.42$ & $1.6$\\
         %Clustering & $2.41\pm 0.94$ & $2.62\pm 1.18$ & $-$ & $-$ \\
         \hline
    \end{tabular}
    \caption{The mass measurements performed in this paper. This table summarizes the discussion of Section~\ref{sec:discussion} and the measurements presented in Figure~\ref{fig:mass} for our LRG samples (\emph{all} and \emph{high-mass}). The quoted splashback radii are in comoving coordinates. The abundance-matching measurements are provided without error bars as we have not modeled the selection function of our LRGs. 
    Most measurements and conversions between $M_\text{200m}$ and $r_\text{sp}$ are computed using a model at the median redshift $\bar{z}=0.44$, identical for both samples (see the end of Section~\ref{sec:fitmass} for details). 
    }
    \label{tab:table_masses}
\end{table}

\subsection{Gravitational constants}
\label{sec:gravity}

In this subsection, we discuss how the combination of the lensing masses and splashback radii measured  above can be used to constrain models of gravity. The principle behind this constraint is the fact that, while General Relativity (GR) predicts that the trajectories of light and massive particles are affected by the same metric perturbation, extended models generally predict a discrepancy between the two. 

In extended models, the equations for the linearized-metric potentials 
\citep[$\Phi$ and $\Psi$, see][]{1980PhRvD..22.1882B} 
can be connected to the background-subtracted matter density $\rho(\bm{x})$ through the following equations \citep{2008JCAP...04..013A, 2008PhRvD..78b4015B, 2010PhRvD..81j4023P},
\begin{align}
    \nabla^2 (\Phi + \Psi) = 8 \pi G  \Sigma(x) \rho(x),  
    \label{eq:Poisson}
    \\
    \nabla^2 \Phi = 4 \pi G \mu(x) \rho(x).
    \label{eq:Poisson2}
\end{align}
In the expressions above, the functions $\mu$ and $\Sigma$, also known as  $G_\text{matter}/G$ and $G_\text{light}/G$ can be in principle a function of space and time (collectively indicated by $x$). We stress that the symbol $\Sigma$, previously used to refer to projected three-dimensional distributions ($\Sigma_\text{g}, \Sigma_\text{m}$), has a different use in this context. These equations are expressed in terms of $\Phi$ and $\Phi + \Psi$ because the trajectories of particles are affected by the first, while the deflection of light is governed by the second. In the presence of only non-relativistic matter,  Einstein's equations in GR reduce to $\Phi=\Psi$ and we have $\Sigma = \mu = 1$. 

The same type of deviation from GR can also be captured in the post-Newtonian parametrization by a multiplicative factor $\gamma$ between the two potentials: $\Psi = \gamma\Phi$. If $\mu, \Sigma$, and $\gamma$ are all constants, the three are trivially related:
\begin{equation}
    \frac{\mu}{\Sigma} = \frac{1+\gamma}{2}.
\end{equation}
Under this same assumption, the ratio between the masses measured through lensing and the mass measured through the dynamics of test particles (e.g., faint galaxies or stars) can be used to constrain these parameters and the literature contains multiple results concerning these extended models. Solar System experiments have constrained $\gamma$ to be consistent with its GR value ($\gamma=1$) up to $5$ significant digits  \citep{2003Natur.425..374B}, but the current measurements at larger scales are substantially less precise. For kpc-sized objects (galaxy-scale), stellar kinematics have been combined with solid lensing measurements to obtain $10$ percent constraints  \citep{Bolton:2006yz, 2018Sci...360.1342C}, while large-scale measurements ($\sim 10-100$ Mpc) can be obtained by combining cosmic shear and redshift space distortion measurements to achieve a similar precision \citep[see, e.g.,][]{2013MNRAS.429.2249S, 2018MNRAS.474.4894J}. As for the scales considered in this paper, a precision of about $30$ percent can be obtained by combining lensing masses with either the kinematics of galaxies inside fully collapsed cluster haloes \citep{2016JCAP...04..023P} or the distribution of hot X-ray emitting gas \citep{2015MNRAS.452.1171W}. However, in this case, the effects of the required assumptions (e.g., spherical symmetry and hydrostatic equilibrium for the gas) are harder to capture. In all cases, no deviation from GR has been measured.

As an example of the power of the measurements presented in Section~\ref{sec:fit}, we present here their implication for beyond-GR effects. On one hand, our lensing signal is a measurement of the amplitude $M_\text{200m, L}$ of the lensing matter density $\rho_L = \rho \Sigma$. On the other hand, the splashback radius $r_\text{sp}$ depends on the amplitude of $ \rho_L \times \mu/\Sigma$ and it is related to the splashback mass $M_\text{200m, sp}$. Therefore, we focus on the ratio of these two amplitudes measured in the high-mass sample:
\begin{align}
    \frac{\mu}{\Sigma} = \frac{M_\text{200m, L}}{M_\text{200m, sp}} = 0.8 \pm 0.4 && \Leftrightarrow && \gamma = 0.6\pm 0.8.
\end{align}

In high-density regions such as the Solar System, the expectation $\gamma = 1$ must be recovered with high precision. Hence, alternative theories of gravity commonly predict scale- and density-dependent effects, which cannot be captured through constant values of $\mu$ and $\Sigma$. Because $r_\text{sp}$ marks a sharp density transition around massive objects, it is more suited to test these complicated dependencies. To provide an example of the constraints possible under this second, more complex, interpretation, we followed \cite{Contigiani_2019} to convert the effects of an additional scale-dependent force (also known as a fifth force) on the location of the splashback radius $r_\text{sp}$. In particular, the model we employed is an extension of self-similar spherical collapse models and neglects any non-isotropic effects, e.g. those introduced by miscentering and halo ellipticity. 

In the context of symmetron gravity \citep{2011PhRvD..84j3521H}, the change in $r_\text{sp}$ introduced by the fifth force is obtained by integrating the trajectories of test particles in the presence or absence of this force. In total, the theory considered has three parameters: 1) $\lambda_0/R(t_0)$, the dimensionless vacuum Compton wavelength of the field that we fix to be $0.05$ times the size of the collapsed object; 2) $z_\text{SSB}$, the redshift corresponding to the moment at which the fifth force is turned on in cosmic history, that we fix at $z_\text{SSB}=1.25$; and 3) $f$, a dimensionless force-strength parameter that is zero in GR. The choices of the fixed values that we imposed are based on physical considerations due to the connection of these gravity models to dark energy while maximizing the impact on splashback. See \cite{Contigiani_2019} for more details.

To match the expectation of the model to observations, we first converted the $M_\text{200m}$ lensing measurement into an expected splashback radius $r_\text{sp, L}$ by reversing the procedure explained at the end of Section~\ref{sec:fitmass} and then compared the measured $r_\text{sp}$ to this value. From the high-mass data, we obtained the following $1\sigma$ constraints:
\begin{align}
\label{eq:resultf}
    \frac{ r_\text{sp, L} - r_\text{sp}}{r_\text{sp, L}} = 0.07 \pm 0.20 && \implies && f < 1.8.
\end{align}

The symmetron theories associated with $z_\text{SSB}\sim 1$ and cluster-sized objects correspond to a coupling mass $M_S$ scale of the order of $10^{-6}$ Planck masses, a region of the parameter space which is still allowed by the solar-system constraints \citep{2011PhRvD..84j3521H} and which has not been explored by other tests of symmetron gravity \citep[see, e.g.,][]{2018PhRvD..98f4019O, 2018LRR....21....1B}. In particular, the upper limit on $f$ produced here directly translates into a constraint on the symmetron field potential of \cite{Contigiani_2019}.\footnote{However, we stress here that this constraint does not have implications for dark energy, as the model considered is not able to drive cosmic acceleration in the absence of a cosmological constant.} In terms of the explicit parameters of the potential, reported here with an additional subscript $s$ for clarity ($M_s, \lambda_s, \mu_s$), we can define the degeneracy line delimiting the boundary of the constraint using the following relations:
\begin{align}
    f \propto \mu_s \lambda_s^{-1} M_s^{-4} && (1+z_\text{ssb})^3 
    \propto M_s^2\mu_s^2.
\end{align}
Therefore, our result shows that we can test the existence of scalar fields with quite weak couplings and directly project these measurements into a broader theory parameter space.

\subsection{Future prospects}
\label{sec:future}

Our results show that the precision of the recovered splashback mass is not comparable to the low uncertainty of the lensing measurements. Because of this, every constraint based on comparing the two is currently limited by the uncertainty of the first. While this paper's focus is not to provide accurate forecasts, we attempt to quantify how we expect these results to improve in the future with larger and deeper samples. In particular, we focus our attention on wide stage-IV surveys such as \emph{Euclid} \citep{laureijs2011euclid} and  Legacy Survey of Space and Time \citep[LSST,][]{2009arXiv0912.0201L}.

First, we investigate how our results can be rescaled. In the process of inferring  $M_\text{200m}$ from $r_\text{sp}$, we find that the relative precision of the former is always a multiple ($3-4$) of the latter. This statement, which we  have verified over a wide range of redshifts ($z \in [0, 1.5]$) and masses ($M_\text{200m} \in [10^{13}, 10^{15}]~\text{M}_\odot$), is a simple consequence of the low slope of the $M_\text{200m}-r_\text{sp}$ relation. Second, we estimate the size of a cluster sample we can obtain and how that translates into an improved errorbar for $r_\text{sp}$. LSST is expected to reach $2.5$ magnitudes deeper than KiDS and to cover an area of the sky $18$ times larger \citep{2009arXiv0912.0201L}. Part of this region is covered by the Galactic plane and will need to be excluded in practice, but the resulting LRG sample will reach up to $z\sim1.2$ and cover a comoving volume about a factor $100$ larger than what is considered in this work. Because the selected LRGs are designed to have a constant comoving density, we can use this estimate to scale the error bars of our galaxy profile measurement. A sample $N=100$ times the size would result in a relative precision in $r_\text{sp}$ of about $2.5$ percent, which translates into a measured $M_\text{200m}$ below $10$ percentage points. This result is obtained by simply re-scaling the error bars of the galaxy profiles by a factor $\sqrt{N} = 10$, but we stress that the effects do not scale linearly for $r_\text{sp}$ due to the slightly skewed posterior of this parameter. While this uncertainty is still larger than what is allowed by lensing measurements, we point out that this method can easily be applied to high-redshift clusters, for which lensing measurements are difficult due to the fewer background sources available (see Figure~\ref{fig:redshift}). 

We note that this simple forecast sidesteps a few issues. Here we consider three of them and discuss their implications and possible solutions. 1) At high redshift, color identification requires additional bands, as the $4000$ \AA~break moves out of the LSST $grizy$ filters. Additional photometry will be required to account for this. 2) Even if we assume that an LRG sample can be constructed, the population of orbiting satellites at high redshift might not necessarily be easy to identify as the red sequence is only beginning to form. Ideally, there is always a color-magnitude galaxy selection that provides a profile compatible with the dark matter profile, but, at this moment, further investigation is required. 3) Finally, with more depth, we also expect fainter satellites to contribute to the galaxy profile signal, but the details of this population for large cluster samples at high redshift are not known. A simple extrapolation of the observed satellite magnitude distribution implies that the number of satellites forming the galaxy distribution signal might be enhanced by an additional factor $10$, reducing the errors in mass to a few percentage points. This, however, is complicated by the fact that different galaxy populations might present profiles inconsistent with the dark matter features \citep{2022arXiv220205277O}. 

In addition to the forecast for the galaxy profiles discussed above, we also expect a measurement of $r_\text{sp}$ with a few percentage point uncertainty directly from the lensing profile \citep{2020MNRAS.499.3534X}. This precision will only be available for relatively low redshifts ($z\sim0.45$), enabling a precise comparison of the dark matter and galaxy profiles. This cross-check can also be used to understand the effects of galaxy evolution in shaping the galaxy phase-space structure \citep{2021arXiv210505914S} and help disentangle the effects of dynamical friction, feedback, and modified models of dark matter \citep{2016JCAP...07..022A, 2020JCAP...02..024B}.

\section{Conclusions}
\label{sec:conclusion}

Accretion connects the mildly non-linear environment of massive haloes to the intrinsic properties of their multi-stream regions. In the last few years, precise measurements of the outer edge of massive dark matter haloes have become feasible  thanks to the introduction of large galaxy samples and a new research field has been opened. 

In this paper, we have used the splashback feature to measure the average dynamical mass of haloes hosting bright KiDS LRGs. To support our result, we have validated this mass measurement using weak lensing masses and a simple abundance-matching argument (see Figure~\ref{fig:mass} and Table~\ref{tab:table_masses}). 

The main achievement that we want to stress here is that these self-consistent measurements are exclusively based on photometric data. In particular, the bright LRG samples used here can be easily matched to simulations, offer a straightforward interpretation, and, in general, are found to be robust against systematic effects in the redshift calibration \citep{2021arXiv210106010B}. This is in contrast to other dynamical mass results presented in the literature: such measurements are based on expensive spectroscopic data \citep[see, e.g., ][]{2016ApJ...819...63R} and are found to produce masses higher than lensing estimates  \citep{2020MNRAS.497.4684H}, an effect which might be due to systematic selection biases afflicting these more accurate measurements \citep{2015MNRAS.449.1897O}. 

Because the relation between $r_\text{sp}$ and halo mass depends on cosmology, this measurement naturally provides a constraint on structure formation.
In this work, we have shown how the combination of splashback and lensing masses has the ability to constrain deviations from GR and the presence of fifth forces (see Section~\ref{sec:gravity}). 

Although the precision of the splashback measurement is relatively low with current data, trends with redshift, mass, and galaxy properties are expected to be informative in the future \citep{2020MNRAS.499.3534X, 2021arXiv210505914S}. Next-generation data will enable new studies of the physics behind galaxy formation \citep{2020arXiv200811663A}, as well as the large-scale environment of massive haloes \citep{2021MNRAS.tmp.1404C}.
As mentioned in Section~\ref{sec:future}, stage IV surveys will substantially advance these new research goals. In particular, we have shown that  splashback masses scale purely with survey volume, unlike lensing. This implies that this technique is uniquely positioned to provide accurate high-redshift masses.

\section*{Acknowledgements}

OC is supported by a de Sitter Fellowship of the Netherlands Organization for Scientific Research (NWO) and by the Natural Sciences and Engineering Research Council of Canada (NSERC). HH, MCF, and MV acknowledge support from the Vici grant No. 639.043.512 financed by the Netherlands Organisation for Scientific Research (NWO). AD is supported by a European Research Council Consolidator Grant No. 770935. ZY acknowledges support from the Max Planck Society and the Alexander von Humboldt Foundation in the framework of the Max Planck-Humboldt Research Award endowed by the Federal Ministry of Education and Research (Germany). CS acknowledges support from the Agencia Nacional de Investigaci\'on y Desarrollo (ANID) through FONDECYT grant no.\ 11191125. All authors contributed to the development and writing of this paper. The authorship list is given in two alphabetical groups: two lead authors (OC, HH) and a list of authors who made a significant contribution to either the data products or the scientific analysis. 

%%%%%%%%%%%%%%%%%%%%%%%%%%%%%%%%%%%%%%%%%%%%%%%%%%
\section*{Data Availability}

The Kilo-Degree Survey data is available at the following link \url{https://kids.strw.leidenuniv.nl/}. The intermediate data products used for this article will be shared at reasonable request to the corresponding authors.



\bibliographystyle{mnras}
\bibliography{bibliography}

\begin{thebibliography}{}
\makeatletter
\relax
\def\mn@urlcharsother{\let\do\@makeother \do\$\do\&\do\#\do\^\do\_\do\%\do\~}
\def\mn@doi{\begingroup\mn@urlcharsother \@ifnextchar [ {\mn@doi@}
  {\mn@doi@[]}}
\def\mn@doi@[#1]#2{\def\@tempa{#1}\ifx\@tempa\@empty \href
  {http://dx.doi.org/#2} {doi:#2}\else \href {http://dx.doi.org/#2} {#1}\fi
  \endgroup}
\def\mn@eprint#1#2{\mn@eprint@#1:#2::\@nil}
\def\mn@eprint@arXiv#1{\href {http://arxiv.org/abs/#1} {{\tt arXiv:#1}}}
\def\mn@eprint@dblp#1{\href {http://dblp.uni-trier.de/rec/bibtex/#1.xml}
  {dblp:#1}}
\def\mn@eprint@#1:#2:#3:#4\@nil{\def\@tempa {#1}\def\@tempb {#2}\def\@tempc
  {#3}\ifx \@tempc \@empty \let \@tempc \@tempb \let \@tempb \@tempa \fi \ifx
  \@tempb \@empty \def\@tempb {arXiv}\fi \@ifundefined
  {mn@eprint@\@tempb}{\@tempb:\@tempc}{\expandafter \expandafter \csname
  mn@eprint@\@tempb\endcsname \expandafter{\@tempc}}}

\bibitem[\protect\citeauthoryear{Adam et~al.,}{Adam et~al.}{2016}]{Planck2015}
Adam R.,  et~al., 2016, \mn@doi [Astronomy & Astrophysics]
  {10.1051/0004-6361/201527101}, 594, A1

\bibitem[\protect\citeauthoryear{Adhikari, Dalal  \& Chamberlain}{Adhikari
  et~al.}{2014}]{Adhikari_2014}
Adhikari S.,  Dalal N.,   Chamberlain R.~T.,  2014, \mn@doi [Journal of
  Cosmology and Astroparticle Physics] {10.1088/1475-7516/2014/11/019}, 2014,
  019–019

\bibitem[\protect\citeauthoryear{{Adhikari}, {Dalal}  \& {Clampitt}}{{Adhikari}
  et~al.}{2016}]{2016JCAP...07..022A}
{Adhikari} S.,  {Dalal} N.,   {Clampitt} J.,  2016, \mn@doi [\jcap]
  {10.1088/1475-7516/2016/07/022}, \href
  {https://ui.adsabs.harvard.edu/abs/2016JCAP...07..022A} {2016, 022}

\bibitem[\protect\citeauthoryear{{Adhikari} et~al.,}{{Adhikari}
  et~al.}{2020}]{2020arXiv200811663A}
{Adhikari} S.,  et~al., 2020, arXiv e-prints, \href
  {https://ui.adsabs.harvard.edu/abs/2020arXiv200811663A} {p. arXiv:2008.11663}

\bibitem[\protect\citeauthoryear{{Albrecht} et~al.,}{{Albrecht}
  et~al.}{2006}]{2006astro.ph..9591A}
{Albrecht} A.,  et~al., 2006, arXiv e-prints, \href
  {https://ui.adsabs.harvard.edu/abs/2006astro.ph..9591A} {pp
  astro--ph/0609591}

\bibitem[\protect\citeauthoryear{{Amendola}, {Kunz}  \& {Sapone}}{{Amendola}
  et~al.}{2008}]{2008JCAP...04..013A}
{Amendola} L.,  {Kunz} M.,   {Sapone} D.,  2008, \mn@doi [\jcap]
  {10.1088/1475-7516/2008/04/013}, \href
  {https://ui.adsabs.harvard.edu/abs/2008JCAP...04..013A} {2008, 013}

\bibitem[\protect\citeauthoryear{{Arnouts} \& {Ilbert}}{{Arnouts} \&
  {Ilbert}}{2011}]{2011ascl.soft08009A}
{Arnouts} S.,  {Ilbert} O.,  2011, {LePHARE: Photometric Analysis for Redshift
  Estimate} (\mn@eprint {ascl} {1108.009})

\bibitem[\protect\citeauthoryear{{Banerjee}, {Adhikari}, {Dalal}, {More}  \&
  {Kravtsov}}{{Banerjee} et~al.}{2020}]{2020JCAP...02..024B}
{Banerjee} A.,  {Adhikari} S.,  {Dalal} N.,  {More} S.,   {Kravtsov} A.,  2020,
  \mn@doi [\jcap] {10.1088/1475-7516/2020/02/024}, \href
  {https://ui.adsabs.harvard.edu/abs/2020JCAP...02..024B} {2020, 024}

\bibitem[\protect\citeauthoryear{{Bardeen}}{{Bardeen}}{1980}]{1980PhRvD..22.1882B}
{Bardeen} J.~M.,  1980, \mn@doi [\prd] {10.1103/PhysRevD.22.1882}, \href
  {https://ui.adsabs.harvard.edu/abs/1980PhRvD..22.1882B} {22, 1882}

\bibitem[\protect\citeauthoryear{{Baxter} et~al.,}{{Baxter}
  et~al.}{2017}]{2017ApJ...841...18B}
{Baxter} E.,  et~al., 2017, \mn@doi [\apj] {10.3847/1538-4357/aa6ff0}, \href
  {https://ui.adsabs.harvard.edu/abs/2017ApJ...841...18B} {841, 18}

\bibitem[\protect\citeauthoryear{{Bertotti}, {Iess}  \& {Tortora}}{{Bertotti}
  et~al.}{2003}]{2003Natur.425..374B}
{Bertotti} B.,  {Iess} L.,   {Tortora} P.,  2003, \mn@doi [\nat]
  {10.1038/nature01997}, \href
  {https://ui.adsabs.harvard.edu/abs/2003Natur.425..374B} {425, 374}

\bibitem[\protect\citeauthoryear{{Bertschinger} \& {Zukin}}{{Bertschinger} \&
  {Zukin}}{2008}]{2008PhRvD..78b4015B}
{Bertschinger} E.,  {Zukin} P.,  2008, \mn@doi [\prd]
  {10.1103/PhysRevD.78.024015}, \href
  {https://ui.adsabs.harvard.edu/abs/2008PhRvD..78b4015B} {78, 024015}

\bibitem[\protect\citeauthoryear{{Bhattacharya}, {Habib}, {Heitmann}  \&
  {Vikhlinin}}{{Bhattacharya} et~al.}{2013}]{2013ApJ...766...32B}
{Bhattacharya} S.,  {Habib} S.,  {Heitmann} K.,   {Vikhlinin} A.,  2013,
  \mn@doi [\apj] {10.1088/0004-637X/766/1/32}, \href
  {https://ui.adsabs.harvard.edu/abs/2013ApJ...766...32B} {766, 32}

\bibitem[\protect\citeauthoryear{{Bilicki} et~al.,}{{Bilicki}
  et~al.}{2021}]{2021arXiv210106010B}
{Bilicki} M.,  et~al., 2021, arXiv e-prints, \href
  {https://ui.adsabs.harvard.edu/abs/2021arXiv210106010B} {p. arXiv:2101.06010}

\bibitem[\protect\citeauthoryear{Bolton, Rappaport  \& Burles}{Bolton
  et~al.}{2006}]{Bolton:2006yz}
Bolton A.~S.,  Rappaport S.,   Burles S.,  2006, \mn@doi [Phys. Rev. D]
  {10.1103/PhysRevD.74.061501}, 74, 061501

\bibitem[\protect\citeauthoryear{{Burrage} \& {Sakstein}}{{Burrage} \&
  {Sakstein}}{2018}]{2018LRR....21....1B}
{Burrage} C.,  {Sakstein} J.,  2018, \mn@doi [Living Reviews in Relativity]
  {10.1007/s41114-018-0011-x}, \href
  {https://ui.adsabs.harvard.edu/abs/2018LRR....21....1B} {21, 1}

\bibitem[\protect\citeauthoryear{Busch \& White}{Busch \&
  White}{2017}]{Busch_2017}
Busch P.,  White S. D.~M.,  2017, \mn@doi [Monthly Notices of the Royal
  Astronomical Society] {10.1093/mnras/stx1584}, 470, 4767–4781

\bibitem[\protect\citeauthoryear{{Chang} et~al.,}{{Chang}
  et~al.}{2018}]{2018ApJ...864...83C}
{Chang} C.,  et~al., 2018, \mn@doi [\apj] {10.3847/1538-4357/aad5e7}, \href
  {https://ui.adsabs.harvard.edu/abs/2018ApJ...864...83C} {864, 83}

\bibitem[\protect\citeauthoryear{{Collett} et~al.,}{{Collett}
  et~al.}{2018}]{2018Sci...360.1342C}
{Collett} T.~E.,  et~al., 2018, \mn@doi [Science] {10.1126/science.aao2469},
  \href {https://ui.adsabs.harvard.edu/abs/2018Sci...360.1342C} {360, 1342}

\bibitem[\protect\citeauthoryear{Contigiani, Vardanyan  \&
  Silvestri}{Contigiani et~al.}{2019a}]{Contigiani_2019}
Contigiani O.,  Vardanyan V.,   Silvestri A.,  2019a, \mn@doi [Physical Review
  D] {10.1103/physrevd.99.064030}, 99

\bibitem[\protect\citeauthoryear{{Contigiani}, {Hoekstra}  \&
  {Bah{\'e}}}{{Contigiani} et~al.}{2019b}]{2019MNRAS.485..408C}
{Contigiani} O.,  {Hoekstra} H.,   {Bah{\'e}} Y.~M.,  2019b, \mn@doi [\mnras]
  {10.1093/mnras/stz404}, \href
  {https://ui.adsabs.harvard.edu/abs/2019MNRAS.485..408C} {485, 408}

\bibitem[\protect\citeauthoryear{{Contigiani}, {Bah{\'e}}  \&
  {Hoekstra}}{{Contigiani} et~al.}{2021}]{2021MNRAS.tmp.1404C}
{Contigiani} O.,  {Bah{\'e}} Y.~M.,   {Hoekstra} H.,  2021, \mn@doi [\mnras]
  {10.1093/mnras/stab1463}, \href
  {https://ui.adsabs.harvard.edu/abs/2021MNRAS.tmp.1404C} {}

\bibitem[\protect\citeauthoryear{Diemer}{Diemer}{2017}]{Diemer_2017}
Diemer B.,  2017, \mn@doi [The Astrophysical Journal Supplement Series]
  {10.3847/1538-4365/aa799c}, 231, 5

\bibitem[\protect\citeauthoryear{Diemer}{Diemer}{2018}]{Diemer:2017bwl}
Diemer B.,  2018, \mn@doi [Astrophys. J. Suppl.] {10.3847/1538-4365/aaee8c},
  239, 35

\bibitem[\protect\citeauthoryear{{Diemer}}{{Diemer}}{2020a}]{2020ApJS..251...17D}
{Diemer} B.,  2020a, \mn@doi [\apjs] {10.3847/1538-4365/abbf51}, \href
  {https://ui.adsabs.harvard.edu/abs/2020ApJS..251...17D} {251, 17}

\bibitem[\protect\citeauthoryear{{Diemer}}{{Diemer}}{2020b}]{2020ApJ...903...87D}
{Diemer} B.,  2020b, \mn@doi [\apj] {10.3847/1538-4357/abbf52}, \href
  {https://ui.adsabs.harvard.edu/abs/2020ApJ...903...87D} {903, 87}

\bibitem[\protect\citeauthoryear{{Diemer} \& {Kravtsov}}{{Diemer} \&
  {Kravtsov}}{2014}]{2014ApJ...789....1D}
{Diemer} B.,  {Kravtsov} A.~V.,  2014, \mn@doi [\apj]
  {10.1088/0004-637X/789/1/1}, \href
  {https://ui.adsabs.harvard.edu/abs/2014ApJ...789....1D} {789, 1}

\bibitem[\protect\citeauthoryear{{Diemer} \& {Kravtsov}}{{Diemer} \&
  {Kravtsov}}{2015}]{2015ApJ...799..108D}
{Diemer} B.,  {Kravtsov} A.~V.,  2015, \mn@doi [\apj]
  {10.1088/0004-637X/799/1/108}, \href
  {https://ui.adsabs.harvard.edu/abs/2015ApJ...799..108D} {799, 108}

\bibitem[\protect\citeauthoryear{{Diemer}, {More}  \& {Kravtsov}}{{Diemer}
  et~al.}{2013}]{2013ApJ...766...25D}
{Diemer} B.,  {More} S.,   {Kravtsov} A.~V.,  2013, \mn@doi [\apj]
  {10.1088/0004-637X/766/1/25}, \href
  {https://ui.adsabs.harvard.edu/abs/2013ApJ...766...25D} {766, 25}

\bibitem[\protect\citeauthoryear{Diemer, Mansfield, Kravtsov  \& More}{Diemer
  et~al.}{2017}]{Diemer_2017b}
Diemer B.,  Mansfield P.,  Kravtsov A.~V.,   More S.,  2017, \mn@doi [The
  Astrophysical Journal] {10.3847/1538-4357/aa79ab}, 843, 140

\bibitem[\protect\citeauthoryear{Dvornik et~al.,}{Dvornik
  et~al.}{2017}]{Dvornik_2017}
Dvornik A.,  et~al., 2017, \mn@doi [Monthly Notices of the Royal Astronomical
  Society] {10.1093/mnras/stx705}, 468, 3251–3265

\bibitem[\protect\citeauthoryear{{Edge}, {Sutherland}, {Kuijken}, {Driver},
  {McMahon}, {Eales}  \& {Emerson}}{{Edge} et~al.}{2013}]{Edge2013}
{Edge} A.,  {Sutherland} W.,  {Kuijken} K.,  {Driver} S.,  {McMahon} R.,
  {Eales} S.,   {Emerson} J.~P.,  2013, The Messenger, \href
  {https://ui.adsabs.harvard.edu/abs/2013Msngr.154...32E} {154, 32}

\bibitem[\protect\citeauthoryear{Einasto}{Einasto}{1965}]{Einasto1965}
Einasto J.,  1965, Trudy Astrofizicheskogo Instituta Alma-Ata, 5, 87

\bibitem[\protect\citeauthoryear{{Erben} et~al.,}{{Erben}
  et~al.}{2005}]{2005AN....326..432E}
{Erben} T.,  et~al., 2005, \mn@doi [Astronomische Nachrichten]
  {10.1002/asna.200510396}, \href
  {https://ui.adsabs.harvard.edu/abs/2005AN....326..432E} {326, 432}

\bibitem[\protect\citeauthoryear{Fenech~Conti, Herbonnet, Hoekstra, Merten,
  Miller  \& Viola}{Fenech~Conti et~al.}{2017}]{Conti:2016gav}
Fenech~Conti I.,  Herbonnet R.,  Hoekstra H.,  Merten J.,  Miller L.,   Viola
  M.,  2017, \mn@doi [Mon. Not. Roy. Astron. Soc.] {10.1093/mnras/stx200}, 467,
  1627

\bibitem[\protect\citeauthoryear{Foreman-Mackey, Hogg, Lang  \&
  Goodman}{Foreman-Mackey et~al.}{2013}]{Foreman-Mackey2013}
Foreman-Mackey D.,  Hogg D.~W.,  Lang D.,   Goodman J.,  2013, \mn@doi
  [Publications of the Astronomical Society of the Pacific] {10.1086/670067},
  125, 306

\bibitem[\protect\citeauthoryear{{Fortuna} et~al.,}{{Fortuna}
  et~al.}{2021}]{MCF2021}
{Fortuna} M.~C.,  et~al., 2021, \mn@doi [\aap] {10.1051/0004-6361/202140706},
  \href {https://ui.adsabs.harvard.edu/abs/2021A&A...654A..76F} {654, A76}

\bibitem[\protect\citeauthoryear{Gao, Navarro, Cole, Frenk, White, Springel,
  Jenkins  \& Neto}{Gao et~al.}{2008}]{Gao2008}
Gao L.,  Navarro J.~F.,  Cole S.,  Frenk C.~S.,  White S. D.~M.,  Springel V.,
  Jenkins A.,   Neto A.~F.,  2008, \mn@doi [Monthly Notices of the Royal
  Astronomical Society] {10.1111/j.1365-2966.2008.13277.x}, 387, 536

\bibitem[\protect\citeauthoryear{{Genel} et~al.,}{{Genel}
  et~al.}{2014}]{2014MNRAS.445..175G}
{Genel} S.,  et~al., 2014, \mn@doi [\mnras] {10.1093/mnras/stu1654}, \href
  {https://ui.adsabs.harvard.edu/abs/2014MNRAS.445..175G} {445, 175}

\bibitem[\protect\citeauthoryear{{Gill}, {Knebe}  \& {Gibson}}{{Gill}
  et~al.}{2005}]{2005MNRAS.356.1327G}
{Gill} S. P.~D.,  {Knebe} A.,   {Gibson} B.~K.,  2005, \mn@doi [\mnras]
  {10.1111/j.1365-2966.2004.08562.x}, \href
  {https://ui.adsabs.harvard.edu/abs/2005MNRAS.356.1327G} {356, 1327}

\bibitem[\protect\citeauthoryear{{Gladders} \& {Yee}}{{Gladders} \&
  {Yee}}{2000}]{2000AJ....120.2148G}
{Gladders} M.~D.,  {Yee} H.~K.~C.,  2000, \mn@doi [\aj] {10.1086/301557}, \href
  {https://ui.adsabs.harvard.edu/abs/2000AJ....120.2148G} {120, 2148}

\bibitem[\protect\citeauthoryear{{Herbonnet} et~al.,}{{Herbonnet}
  et~al.}{2020}]{2020MNRAS.497.4684H}
{Herbonnet} R.,  et~al., 2020, \mn@doi [\mnras] {10.1093/mnras/staa2303}, \href
  {https://ui.adsabs.harvard.edu/abs/2020MNRAS.497.4684H} {497, 4684}

\bibitem[\protect\citeauthoryear{{Hinterbichler}, {Khoury}, {Levy}  \&
  {Matas}}{{Hinterbichler} et~al.}{2011}]{2011PhRvD..84j3521H}
{Hinterbichler} K.,  {Khoury} J.,  {Levy} A.,   {Matas} A.,  2011, \mn@doi
  [\prd] {10.1103/PhysRevD.84.103521}, \href
  {https://ui.adsabs.harvard.edu/abs/2011PhRvD..84j3521H} {84, 103521}

\bibitem[\protect\citeauthoryear{{Jarvis}}{{Jarvis}}{2015}]{2015ascl.soft08007J}
{Jarvis} M.,  2015, {TreeCorr: Two-point correlation functions} (\mn@eprint
  {ascl} {1508.007})

\bibitem[\protect\citeauthoryear{{Jarvis}, {Bernstein}  \& {Jain}}{{Jarvis}
  et~al.}{2004}]{2004MNRAS.352..338J}
{Jarvis} M.,  {Bernstein} G.,   {Jain} B.,  2004, \mn@doi [\mnras]
  {10.1111/j.1365-2966.2004.07926.x}, \href
  {https://ui.adsabs.harvard.edu/abs/2004MNRAS.352..338J} {352, 338}

\bibitem[\protect\citeauthoryear{{Joudaki} et~al.,}{{Joudaki}
  et~al.}{2018}]{2018MNRAS.474.4894J}
{Joudaki} S.,  et~al., 2018, \mn@doi [\mnras] {10.1093/mnras/stx2820}, \href
  {https://ui.adsabs.harvard.edu/abs/2018MNRAS.474.4894J} {474, 4894}

\bibitem[\protect\citeauthoryear{{Kannawadi} et~al.,}{{Kannawadi}
  et~al.}{2019}]{2019A&A...624A..92K}
{Kannawadi} A.,  et~al., 2019, \mn@doi [\aap] {10.1051/0004-6361/201834819},
  \href {https://ui.adsabs.harvard.edu/abs/2019A&A...624A..92K} {624, A92}

\bibitem[\protect\citeauthoryear{{Kuijken} et~al.,}{{Kuijken}
  et~al.}{2019}]{Kuijken2019DR4}
{Kuijken} K.,  et~al., 2019, \mn@doi [\aap] {10.1051/0004-6361/201834918},
  \href {https://ui.adsabs.harvard.edu/abs/2019A&A...625A...2K} {625, A2}

\bibitem[\protect\citeauthoryear{{LSST Science Collaboration} et~al.,}{{LSST
  Science Collaboration} et~al.}{2009}]{2009arXiv0912.0201L}
{LSST Science Collaboration} et~al., 2009, arXiv e-prints, \href
  {https://ui.adsabs.harvard.edu/abs/2009arXiv0912.0201L} {p. arXiv:0912.0201}

\bibitem[\protect\citeauthoryear{{Larson}, {Tinsley}  \& {Caldwell}}{{Larson}
  et~al.}{1980}]{1980ApJ...237..692L}
{Larson} R.~B.,  {Tinsley} B.~M.,   {Caldwell} C.~N.,  1980, \mn@doi [\apj]
  {10.1086/157917}, \href
  {https://ui.adsabs.harvard.edu/abs/1980ApJ...237..692L} {237, 692}

\bibitem[\protect\citeauthoryear{Laureijs et~al.,}{Laureijs
  et~al.}{2011}]{laureijs2011euclid}
Laureijs R.,  et~al., 2011, Euclid Definition Study Report (\mn@eprint {arXiv}
  {1110.3193})

\bibitem[\protect\citeauthoryear{{Mahajan}, {Mamon}  \&
  {Raychaudhury}}{{Mahajan} et~al.}{2011}]{2011MNRAS.416.2882M}
{Mahajan} S.,  {Mamon} G.~A.,   {Raychaudhury} S.,  2011, \mn@doi [\mnras]
  {10.1111/j.1365-2966.2011.19236.x}, \href
  {https://ui.adsabs.harvard.edu/abs/2011MNRAS.416.2882M} {416, 2882}

\bibitem[\protect\citeauthoryear{{Mancone} \& {Gonzalez}}{{Mancone} \&
  {Gonzalez}}{2012}]{2012PASP..124..606M}
{Mancone} C.~L.,  {Gonzalez} A.~H.,  2012, \mn@doi [\pasp] {10.1086/666502},
  \href {https://ui.adsabs.harvard.edu/abs/2012PASP..124..606M} {124, 606}

\bibitem[\protect\citeauthoryear{{Mandelbaum}, {Seljak}, {Kauffmann}, {Hirata}
  \& {Brinkmann}}{{Mandelbaum} et~al.}{2006}]{2006MNRAS.368..715M}
{Mandelbaum} R.,  {Seljak} U.,  {Kauffmann} G.,  {Hirata} C.~M.,   {Brinkmann}
  J.,  2006, \mn@doi [\mnras] {10.1111/j.1365-2966.2006.10156.x}, \href
  {https://ui.adsabs.harvard.edu/abs/2006MNRAS.368..715M} {368, 715}

\bibitem[\protect\citeauthoryear{{McFarland}, {Verdoes-Kleijn}, {Sikkema},
  {Helmich}, {Boxhoorn}  \& {Valentijn}}{{McFarland}
  et~al.}{2013}]{2013ExA....35...45M}
{McFarland} J.~P.,  {Verdoes-Kleijn} G.,  {Sikkema} G.,  {Helmich} E.~M.,
  {Boxhoorn} D.~R.,   {Valentijn} E.~A.,  2013, \mn@doi [Experimental
  Astronomy] {10.1007/s10686-011-9266-x}, \href
  {https://ui.adsabs.harvard.edu/abs/2013ExA....35...45M} {35, 45}

\bibitem[\protect\citeauthoryear{{Miller} et~al.,}{{Miller}
  et~al.}{2013}]{2013MNRAS.429.2858M}
{Miller} L.,  et~al., 2013, \mn@doi [\mnras] {10.1093/mnras/sts454}, \href
  {https://ui.adsabs.harvard.edu/abs/2013MNRAS.429.2858M} {429, 2858}

\bibitem[\protect\citeauthoryear{{Moore}, {Katz}, {Lake}, {Dressler}  \&
  {Oemler}}{{Moore} et~al.}{1996}]{1996Natur.379..613M}
{Moore} B.,  {Katz} N.,  {Lake} G.,  {Dressler} A.,   {Oemler} A.,  1996,
  \mn@doi [\nat] {10.1038/379613a0}, \href
  {https://ui.adsabs.harvard.edu/abs/1996Natur.379..613M} {379, 613}

\bibitem[\protect\citeauthoryear{More, Diemer  \& Kravtsov}{More
  et~al.}{2015}]{More_2015}
More S.,  Diemer B.,   Kravtsov A.~V.,  2015, \mn@doi [The Astrophysical
  Journal] {10.1088/0004-637x/810/1/36}, 810, 36

\bibitem[\protect\citeauthoryear{{More} et~al.,}{{More}
  et~al.}{2016}]{2016ApJ...825...39M}
{More} S.,  et~al., 2016, \mn@doi [\apj] {10.3847/0004-637X/825/1/39}, \href
  {https://ui.adsabs.harvard.edu/abs/2016ApJ...825...39M} {825, 39}

\bibitem[\protect\citeauthoryear{Navarro, Frenk  \& White}{Navarro
  et~al.}{1996}]{Navarro1996}
Navarro J.~F.,  Frenk C.~S.,   White S. D.~M.,  1996, \mn@doi [The
  Astrophysical Journal] {10.1086/177173}, 462, 563

\bibitem[\protect\citeauthoryear{Navarro, Frenk  \& White}{Navarro
  et~al.}{1997}]{Navarro1997}
Navarro J.~F.,  Frenk C.~S.,   White S. D.~M.,  1997, \mn@doi [The
  Astrophysical Journal] {10.1086/304888}, 490, 493

\bibitem[\protect\citeauthoryear{{O'Hare} \& {Burrage}}{{O'Hare} \&
  {Burrage}}{2018}]{2018PhRvD..98f4019O}
{O'Hare} C. A.~J.,  {Burrage} C.,  2018, \mn@doi [\prd]
  {10.1103/PhysRevD.98.064019}, \href
  {https://ui.adsabs.harvard.edu/abs/2018PhRvD..98f4019O} {98, 064019}

\bibitem[\protect\citeauthoryear{{O'Neil}, {Barnes}, {Vogelsberger}  \&
  {Diemer}}{{O'Neil} et~al.}{2021}]{2021MNRAS.504.4649O}
{O'Neil} S.,  {Barnes} D.~J.,  {Vogelsberger} M.,   {Diemer} B.,  2021, \mn@doi
  [\mnras] {10.1093/mnras/stab1221}, \href
  {https://ui.adsabs.harvard.edu/abs/2021MNRAS.504.4649O} {504, 4649}

\bibitem[\protect\citeauthoryear{{O'Neil}, {Borrow}, {Vogelsberger}  \&
  {Diemer}}{{O'Neil} et~al.}{2022}]{2022arXiv220205277O}
{O'Neil} S.,  {Borrow} J.,  {Vogelsberger} M.,   {Diemer} B.,  2022, arXiv
  e-prints, \href {https://ui.adsabs.harvard.edu/abs/2022arXiv220205277O} {p.
  arXiv:2202.05277}

\bibitem[\protect\citeauthoryear{{Old} et~al.,}{{Old}
  et~al.}{2015}]{2015MNRAS.449.1897O}
{Old} L.,  et~al., 2015, \mn@doi [\mnras] {10.1093/mnras/stv421}, \href
  {https://ui.adsabs.harvard.edu/abs/2015MNRAS.449.1897O} {449, 1897}

\bibitem[\protect\citeauthoryear{{Pizzuti} et~al.,}{{Pizzuti}
  et~al.}{2016}]{2016JCAP...04..023P}
{Pizzuti} L.,  et~al., 2016, \mn@doi [\jcap] {10.1088/1475-7516/2016/04/023},
  \href {https://ui.adsabs.harvard.edu/abs/2016JCAP...04..023P} {2016, 023}

\bibitem[\protect\citeauthoryear{{Pogosian}, {Silvestri}, {Koyama}  \&
  {Zhao}}{{Pogosian} et~al.}{2010}]{2010PhRvD..81j4023P}
{Pogosian} L.,  {Silvestri} A.,  {Koyama} K.,   {Zhao} G.-B.,  2010, \mn@doi
  [\prd] {10.1103/PhysRevD.81.104023}, \href
  {https://ui.adsabs.harvard.edu/abs/2010PhRvD..81j4023P} {81, 104023}

\bibitem[\protect\citeauthoryear{{Press} \& {Schechter}}{{Press} \&
  {Schechter}}{1974}]{1974ApJ...187..425P}
{Press} W.~H.,  {Schechter} P.,  1974, \mn@doi [\apj] {10.1086/152650}, \href
  {https://ui.adsabs.harvard.edu/abs/1974ApJ...187..425P} {187, 425}

\bibitem[\protect\citeauthoryear{Price-Whelan et~al.}{Price-Whelan
  et~al.}{2018}]{Price-Whelan:2018hus}
Price-Whelan A.~M.,  et~al., 2018, \mn@doi [Astron. J.]
  {10.3847/1538-3881/aabc4f}, 156, 123

\bibitem[\protect\citeauthoryear{{Rines}, {Geller}, {Diaferio}  \&
  {Hwang}}{{Rines} et~al.}{2016}]{2016ApJ...819...63R}
{Rines} K.~J.,  {Geller} M.~J.,  {Diaferio} A.,   {Hwang} H.~S.,  2016, \mn@doi
  [\apj] {10.3847/0004-637X/819/1/63}, \href
  {https://ui.adsabs.harvard.edu/abs/2016ApJ...819...63R} {819, 63}

\bibitem[\protect\citeauthoryear{{Rozo} et~al.,}{{Rozo}
  et~al.}{2016}]{2016MNRAS.461.1431R}
{Rozo} E.,  et~al., 2016, \mn@doi [\mnras] {10.1093/mnras/stw1281}, \href
  {https://ui.adsabs.harvard.edu/abs/2016MNRAS.461.1431R} {461, 1431}

\bibitem[\protect\citeauthoryear{{Schaye} et~al.,}{{Schaye}
  et~al.}{2010}]{2010MNRAS.402.1536S}
{Schaye} J.,  et~al., 2010, \mn@doi [\mnras]
  {10.1111/j.1365-2966.2009.16029.x}, \href
  {https://ui.adsabs.harvard.edu/abs/2010MNRAS.402.1536S} {402, 1536}

\bibitem[\protect\citeauthoryear{{Sereno}, {Fedeli}  \& {Moscardini}}{{Sereno}
  et~al.}{2016}]{2016JCAP...01..042S}
{Sereno} M.,  {Fedeli} C.,   {Moscardini} L.,  2016, \mn@doi [\jcap]
  {10.1088/1475-7516/2016/01/042}, \href
  {https://ui.adsabs.harvard.edu/abs/2016JCAP...01..042S} {2016, 042}

\bibitem[\protect\citeauthoryear{Shin et~al.,}{Shin et~al.}{2019}]{Shin_2019}
Shin T.,  et~al., 2019, \mn@doi [Monthly Notices of the Royal Astronomical
  Society] {10.1093/mnras/stz1434}, 487, 2900–2918

\bibitem[\protect\citeauthoryear{{Shin} et~al.,}{{Shin}
  et~al.}{2021}]{2021arXiv210505914S}
{Shin} T.,  et~al., 2021, arXiv e-prints, \href
  {https://ui.adsabs.harvard.edu/abs/2021arXiv210505914S} {p. arXiv:2105.05914}

\bibitem[\protect\citeauthoryear{{Simpson} et~al.,}{{Simpson}
  et~al.}{2013}]{2013MNRAS.429.2249S}
{Simpson} F.,  et~al., 2013, \mn@doi [\mnras] {10.1093/mnras/sts493}, \href
  {https://ui.adsabs.harvard.edu/abs/2013MNRAS.429.2249S} {429, 2249}

\bibitem[\protect\citeauthoryear{{Somerville}, {Hopkins}, {Cox}, {Robertson}
  \& {Hernquist}}{{Somerville} et~al.}{2008}]{2008MNRAS.391..481S}
{Somerville} R.~S.,  {Hopkins} P.~F.,  {Cox} T.~J.,  {Robertson} B.~E.,
  {Hernquist} L.,  2008, \mn@doi [\mnras] {10.1111/j.1365-2966.2008.13805.x},
  \href {https://ui.adsabs.harvard.edu/abs/2008MNRAS.391..481S} {391, 481}

\bibitem[\protect\citeauthoryear{{Strateva} et~al.,}{{Strateva}
  et~al.}{2001}]{2001AJ....122.1861S}
{Strateva} I.,  et~al., 2001, \mn@doi [\aj] {10.1086/323301}, \href
  {https://ui.adsabs.harvard.edu/abs/2001AJ....122.1861S} {122, 1861}

\bibitem[\protect\citeauthoryear{{Tinker}, {Kravtsov}, {Klypin}, {Abazajian},
  {Warren}, {Yepes}, {Gottl{\"o}ber}  \& {Holz}}{{Tinker}
  et~al.}{2008}]{2008ApJ...688..709T}
{Tinker} J.,  {Kravtsov} A.~V.,  {Klypin} A.,  {Abazajian} K.,  {Warren} M.,
  {Yepes} G.,  {Gottl{\"o}ber} S.,   {Holz} D.~E.,  2008, \mn@doi [\apj]
  {10.1086/591439}, \href
  {https://ui.adsabs.harvard.edu/abs/2008ApJ...688..709T} {688, 709}

\bibitem[\protect\citeauthoryear{{Trayford} et~al.,}{{Trayford}
  et~al.}{2015}]{2015MNRAS.452.2879T}
{Trayford} J.~W.,  et~al., 2015, \mn@doi [\mnras] {10.1093/mnras/stv1461},
  \href {https://ui.adsabs.harvard.edu/abs/2015MNRAS.452.2879T} {452, 2879}

\bibitem[\protect\citeauthoryear{{Umetsu} \& {Diemer}}{{Umetsu} \&
  {Diemer}}{2017}]{2017ApJ...836..231U}
{Umetsu} K.,  {Diemer} B.,  2017, \mn@doi [\apj] {10.3847/1538-4357/aa5c90},
  \href {https://ui.adsabs.harvard.edu/abs/2017ApJ...836..231U} {836, 231}

\bibitem[\protect\citeauthoryear{{Vakili} et~al.,}{{Vakili}
  et~al.}{2019}]{2019MNRAS.487.3715V}
{Vakili} M.,  et~al., 2019, \mn@doi [\mnras] {10.1093/mnras/stz1249}, \href
  {https://ui.adsabs.harvard.edu/abs/2019MNRAS.487.3715V} {487, 3715}

\bibitem[\protect\citeauthoryear{{Vakili} et~al.,}{{Vakili}
  et~al.}{2020}]{2020arXiv200813154V}
{Vakili} M.,  et~al., 2020, arXiv e-prints, \href
  {https://ui.adsabs.harvard.edu/abs/2020arXiv200813154V} {p. arXiv:2008.13154}

\bibitem[\protect\citeauthoryear{{Viola} et~al.,}{{Viola}
  et~al.}{2015}]{2015MNRAS.452.3529V}
{Viola} M.,  et~al., 2015, \mn@doi [\mnras] {10.1093/mnras/stv1447}, \href
  {https://ui.adsabs.harvard.edu/abs/2015MNRAS.452.3529V} {452, 3529}

\bibitem[\protect\citeauthoryear{{Wilcox} et~al.,}{{Wilcox}
  et~al.}{2015}]{2015MNRAS.452.1171W}
{Wilcox} H.,  et~al., 2015, \mn@doi [\mnras] {10.1093/mnras/stv1366}, \href
  {https://ui.adsabs.harvard.edu/abs/2015MNRAS.452.1171W} {452, 1171}

\bibitem[\protect\citeauthoryear{{Xhakaj}, {Diemer}, {Leauthaud}, {Wasserman},
  {Huang}, {Luo}, {Adhikari}  \& {Singh}}{{Xhakaj}
  et~al.}{2020}]{2020MNRAS.499.3534X}
{Xhakaj} E.,  {Diemer} B.,  {Leauthaud} A.,  {Wasserman} A.,  {Huang} S.,
  {Luo} Y.,  {Adhikari} S.,   {Singh} S.,  2020, \mn@doi [\mnras]
  {10.1093/mnras/staa3046}, \href
  {https://ui.adsabs.harvard.edu/abs/2020MNRAS.499.3534X} {499, 3534}

\bibitem[\protect\citeauthoryear{{de Jong}, {Verdoes Kleijn}, {Kuijken}  \&
  {Valentijn}}{{de Jong} et~al.}{2013}]{deJong2013}
{de Jong} J.~T.~A.,  {Verdoes Kleijn} G.~A.,  {Kuijken} K.~H.,   {Valentijn}
  E.~A.,  2013, \mn@doi [Experimental Astronomy] {10.1007/s10686-012-9306-1},
  \href {http://adsabs.harvard.edu/abs/2013ExA....35...25D} {35, 25}

\bibitem[\protect\citeauthoryear{{van Uitert}, {Cacciato}, {Hoekstra}  \&
  {Herbonnet}}{{van Uitert} et~al.}{2015}]{2015A&A...579A..26V}
{van Uitert} E.,  {Cacciato} M.,  {Hoekstra} H.,   {Herbonnet} R.,  2015,
  \mn@doi [\aap] {10.1051/0004-6361/201525834}, \href
  {https://ui.adsabs.harvard.edu/abs/2015A&A...579A..26V} {579, A26}

\bibitem[\protect\citeauthoryear{{van den Bosch}, {Aquino}, {Yang}, {Mo},
  {Pasquali}, {McIntosh}, {Weinmann}  \& {Kang}}{{van den Bosch}
  et~al.}{2008}]{2008MNRAS.387...79V}
{van den Bosch} F.~C.,  {Aquino} D.,  {Yang} X.,  {Mo} H.~J.,  {Pasquali} A.,
  {McIntosh} D.~H.,  {Weinmann} S.~M.,   {Kang} X.,  2008, \mn@doi [\mnras]
  {10.1111/j.1365-2966.2008.13230.x}, \href
  {https://ui.adsabs.harvard.edu/abs/2008MNRAS.387...79V} {387, 79}

\makeatother
\end{thebibliography}

%\begin{thebibliography}{99}
%\end{thebibliography}

%%%%%%%%%%%%%%%%%%%%%%%%%%%%%%%%%%%%%%%%%%%%%%%%%%

% Don't change these lines
\bsp	% typesetting comment
\label{lastpage}
\end{document}